\documentclass[aps, prd, amsmath, floats, floatfix, twocolumn, nofootinbib, superscriptaddress, showpacs]{revtex4-1}
\usepackage{graphicx,color}
\usepackage[dvipsnames]{xcolor}  % more colors
\usepackage{latexsym}
\usepackage{amsmath,amsfonts,amssymb, amsthm}
\usepackage{physics}     % convenient physics shortcuts
\usepackage{braket}
\usepackage{dsfont}      % for \mathds{1}
\usepackage{varwidth}
\usepackage{hyperref}

\def\bC{{\mathbb C}}       
\def\bR{{\mathbb R}} 

\def\oo{\infty}
\def\H{\mathcal{H}}
\def\A{\mathbf{A}}
\def\E{\mathbf{E}}
\def\P{\mathbf{P}}
\def\R{\mathbf{R}}
\def\bS{\mathbf{S}}
\def\G{\mathbf{G}}
\def\T{\mathbf{T}}
\def\cG{\mathcal{G}}
\def\cJ{\mathcal{J}}
\def\cK{\mathcal{K}}

\def\cR{\mathcal{R}}

\def\Res{\operatorname{Res}}
\def\BS{\mathrm{BS}}

\def\tomega{{\tilde{\omega}}}

\begin{document}

% TITLE

\title{Ground state for a massive scalar field in BTZ spacetime \\ with Robin boundary conditions}

% AUTHORS

\author{Francesco Bussola}
\email{francesco.bussola01@ateneopv.it}
\affiliation{Dipartimento di Fisica, Universit\`a degli Studi di Pavia, Via Bassi, 6, 27100 Pavia, Italy}
\affiliation{Istituto Nazionale di Fisica Nucleare -- Sezione di Pavia, Via Bassi, 6, 27100 Pavia, Italy}

\author{Claudio Dappiaggi}
\email{claudio.dappiaggi@unipv.it}
\affiliation{Dipartimento di Fisica, Universit\`a degli Studi di Pavia, Via Bassi, 6, 27100 Pavia, Italy}
\affiliation{Istituto Nazionale di Fisica Nucleare -- Sezione di Pavia, Via Bassi, 6, 27100 Pavia, Italy}

\author{Hugo R. C. Ferreira}
\email{hugo.ferreira@pv.infn.it}
\affiliation{Istituto Nazionale di Fisica Nucleare -- Sezione di Pavia, Via Bassi, 6, 27100 Pavia, Italy}

\author{Igor Khavkine}
\email{igor.khavkine@unimi.it}
\affiliation{Dipartimento di Matematica, Universit\`a di Milano, Via Cesare Saldini, 50, 20133 Milano, Italy}
\affiliation{Istituto Nazionale di Fisica Nucleare -- Sezione di Milano, Via Celoria, 16, 20133 Milan, Italy}

% DATE

\date{\today}

\begin{abstract}
We consider a real, massive scalar field in BTZ spacetime, a 2+1-dimensional black hole solution of the Einstein's field equations with a negative cosmological constant. First, we analyze the space of classical solutions in a mode decomposition and we characterize the collection of all admissible boundary conditions of Robin type which can be imposed at infinity. Secondly, we investigate whether, for a given boundary condition, there exists a ground state by constructing explicitly its two-point function. We demonstrate that for a subclass of the boundary conditions it is possible to construct a ground state that locally satisfies the Hadamard property. In all other cases, we show that bound state mode solutions exist and, therefore, such construction is not possible.
\vspace*{7ex}
\end{abstract}

\maketitle

% INTRODUCTION

\section{Introduction}

Quantum field theory on curved backgrounds is a well-established branch
of theoretical and mathematical physics which allows to study matter
systems in the presence of a non vanishing gravitational field (for a
recent review see Ref.~\cite{Benini:2013fia}). In this framework it is
always assumed both that no proper quantum gravitational effect has to
be accounted for and that the backreaction in the Einstein's equations
is negligible. 

Although this entails that the geometry of the spacetime is fixed, it is
not at all necessary to consider metrics which are small perturbations
over a flat background. Actually quantum field theory in the presence of
a strong gravitational field, {\it e.g.}, a black hole, is of great
interest since one can unveil some novel phenomena, the most famous
example being Hawking radiation \cite{Hawking:1974sw}, which have no
counterpart on Minkowski spacetime. 

For this reason a lot of attention has always been given to the
investigation and to the formulation of quantized free field theories on
black hole spacetimes. Especially under the additional assumption of
spherical symmetry, many results have been obtained leading to an almost
complete understanding of these matter systems both at the structural
and at the physical level
\cite{Christensen:1976vb,Candelas:1980zt,Howard:1984qp,Fredenhagen:1989kr,
Anderson:1990jh,Anderson:1993if,Anderson:1993if,Anderson:1994hg}.

Much more complicated is the scenario when the underlying black hole
solution of the Einstein's equations is rotating, hence only
axisymmetric, the most notable example being Kerr spacetime. In this
case even the analysis of free quantum theories is more elusive and
simple questions like the construction of a ground state in the region
outside the event horizon are difficult to answer \cite{Kay:1988mu}. As
a byproduct, the computation of renormalized physical observables has
been a daunting task
\cite{Frolov:1982pi,Ottewill:2000qh,Duffy:2005mz,Ferreira:2014ina} and
only very recently a promising renormalization scheme has been applied
with success \cite{Levi:2015eea,Levi:2016exv}. The main reason for this
quandary can be ascribed to a peculiarity of such rotating geometries,
namely the absence of a complete, everywhere timelike Killing field. If
it existed, the latter would allow for the identification of a canonical
and natural choice for the notion of positive frequency, which can be
used in turn to select a distinguished two-point function for the
underlying theory. This defines uniquely and unambiguously a
full-fledged quantum state, dubbed the {\em ground state}, with notable
physical and structural properties \cite{Sahlmann:2000fh}. Among them we
recall in particular that all quantum observables have finite
fluctuations and that, starting from such a state, it is possible to
construct the algebra of all Wick polynomials, including relevant
objects such as the stress-energy tensor \cite{Khavkine:2014mta}.

Therefore, on account of the lack of a global timelike Killing field,
our understanding of quantum field theories in presence of rotating
black holes is not as advanced as one could hope. The main goal of this
paper is to discuss a concrete scenario where most of the problems
mentioned above can be circumvented. We refer to the so-called BTZ
black-hole \cite{Banados:1992wn,Banados:1992gq}, a solution of the
(2+1)-dimensional Einstein's equations with a negative cosmological
constant. This geometry possesses some rather peculiar features. On the
one hand, it can be obtained directly from the anti-de Sitter (AdS)
metric with an appropriate identification of boundaries---see
\cite{Banados:1992gq}---, hence locally it is a region of constant
curvature. On the other hand, the BTZ solution is both stationary and
axisymmetric and possesses an inner and an outer horizon, as well as two
canonical Killing fields, say $\partial_t$ and $\partial_\phi$,
associated to these symmetries. In addition, contrary to what happens
in the Kerr spacetime, although none of these vector fields is
everywhere timelike, there exists a suitable linear combination which
enjoys such property everywhere in the region exterior of the black hole.

This feature prompts the possibility of analyzing free field
theories in the BTZ background constructing an associated ground state.
In this paper we will address this issue thoroughly for the case
of a real, massive Klein-Gordon field obeying Robin boundary conditions
at conformal infinity. In this respect, we generalize and complement the results
of Ref.~\cite{Lifschytz:1993eb}, which considers a massless,
conformally coupled scalar field with either Dirichlet or Neumann
boundary conditions at infinity.

Our analysis starts from the construction, via a mode expansion, of the
space of solutions for the underlying equation of motion. This must be
approached delicately, since the underlying spacetime shares both
locally and asymptotically the geometry of anti-de Sitter spacetime. In
particular, this entails that a BTZ black hole spacetime is not globally
hyperbolic, which is tantamount to saying that an acausal spacelike
surface can be at most partially Cauchy (at least some complete timelike
curves will never intersect it) and that the solutions of the equations
of motion for a free field theory cannot be obtained only by imposing
suitable initial data on such a partial Cauchy surface. As a matter of
fact, the existence of a timelike conformal boundary at infinity requires
additional boundary conditions thereon. Here, we follow the same path 
taken in the analysis of a pure AdS spacetime, recently investigated in
\cite{Dappiaggi:2016fwc,Dappiaggi:2017wvj}. This was based on
a careful use of the Sturm-Liouville theory for ordinary differential
equations (ODEs) \cite{Zettl:2005,weidmann} which complements the earlier analyses in \cite{Ishibashi:2003jd,Ishibashi:2004wx,Seggev:2003rp}. By using and extending
similar methods, we will show that also in the BTZ spacetime there
exists a one parameter family of admissible boundary conditions, of
Robin type, depending on the value of the effective squared mass in the
Klein-Gordon equation. From a physical point of view, this guarantees
vanishing energy flux through conformal infinity \cite{Ferreira:2017}.
These conditions are different from the ``transparent boundary
conditions'' used in Ref.~\cite{Steif:1993zv} to compute the
renormalized stress-energy tensor for a massless, conformally coupled
scalar field in the BTZ black hole.

A thorough discussion of this feature is of paramount importance for the
core goal of this work: the construction of the two-point function of a
ground state. In fact, each different boundary condition identifies, for
all practical purposes, a separate dynamical theory. To each of these,
one can compute a distinguished two-point function associated to the ground state defined 
with respect to the timelike Killing field which exists in the region outside the outer horizon. In
the main body of the paper, we not only construct such two-point
functions explicitly, but we also investigate their physical properties.
Most notably, we show that, for a large class of the Robin boundary
conditions, including the Dirichlet one, only positive frequencies
contribute to the mode expansion of the two-point function. Hence, for
each of these admissible boundary conditions, we identify a full-fledged
ground state, which moreover is locally of Hadamard form on account of
some structural results of quantum field theory on curved backgrounds
proven in \cite{Sahlmann:2000fh}. By saying \emph{locally}, we distinguish
from the \emph{global} feature observed in \cite{Dappiaggi:2016fwc,Dappiaggi:2017wvj} 
for a free quantum field theory in the Poincar\'e patch
of anti-de Sitter spacetime: on account of the presence
of the boundary and independently of the chosen boundary condition, the
two-point function is singular not only at those pairs of points
connected by a null geodesic, but also at those which can be reached
after such a geodesic is reflected at the conformal boundary. The
presence of these additional singularities cannot be inferred from the
standard structural properties proven in \cite{Sahlmann:2000fh} and it
requires a more involved mathematical analysis, which is outside the
scope of the present work.  

In addition, we confirm the existence of a rather peculiar feature which
was already observed in the analysis of a real, massive scalar field in
the Poincar\'e patch of anti-de Sitter spacetime \cite{Dappiaggi:2016fwc}. There exists a class
of Robin boundary conditions for which the mode expansion of the
two-point function necessarily includes the contribution of bound state
mode solutions. For these boundary conditions, one cannot
claim that the constructed two-point function is that of a ground
state and, more importantly, that it is of Hadamard form.

The paper will be organized as follows. In
Section~\ref{sec:BTZ_geometry} we review the geometry of a BTZ black
hole, emphasizing in particular the presence of an everywhere timelike
Killing field in the exterior region of the black hole. In
Section~\ref{sec:KGeq}, we analyze the massive Klein-Gordon equation 
with an arbitrary coupling to scalar curvature on this background.
Via a Fourier expansion, the field equation is reduced to an ODE in the radial direction, 
which can be solved explicitly. The solutions are classified in terms of their square 
integrability near the horizon and the conformal infinity, which gives us the range of 
the effective squared mass of the scalar field for which Robin boundary conditions 
have to be imposed at conformal infinity. Finally, in
Section~\ref{Sec:ground_state}, we obtain our main result, namely the
explicit construction for the two-point function of the ground state for
a large class of Robin boundary conditions. Those not in this set are
shown not to possess a ground state, given the presence of bound state
mode solutions. In Section~\ref{sec:conclusions}, we draw our
conclusions. In Appendix~\ref{apx:deltaexpansion} we
discuss how to handle a key technical problem in our construction of the 
two-point function: contrary to what happens when dealing with a scalar field
in a static spacetime, the ODE obtained out of the Fourier analysis
cannot be interpreted as a simple eigenvalue problem with $\omega^2$ as
the spectral parameter, where $\omega$ is the frequency. In fact, the ensuing equation, having also a 
linear dependence in $\omega$, can be read as a so-called quadratic operator pencil. 
In Appendix~\ref{apx:calculation-delta-expansion} we present all the steps of the calculation 
of the two-point function for the ground state, whose results are presented in Section~\ref{Sec:ground_state}. 
We leave some of the most mathematical details for Appendices~\ref{apx:S1}, \ref{apx:S2} 
and \ref{apx:S3}.

Throughout the paper we employ natural units in which $c = G_{\rm N} =
\hbar = 1$ and a metric with signature $({-}{+}{+})$.

% BTZ BLACK HOLE AND 2+1 GEOMETRY

\section{BTZ black hole and 2+1 geometry}
\label{sec:BTZ_geometry}

%%%%%%%%%%%%%%%%

\begin{figure}[t]
\begin{center}
\includegraphics[scale=1]{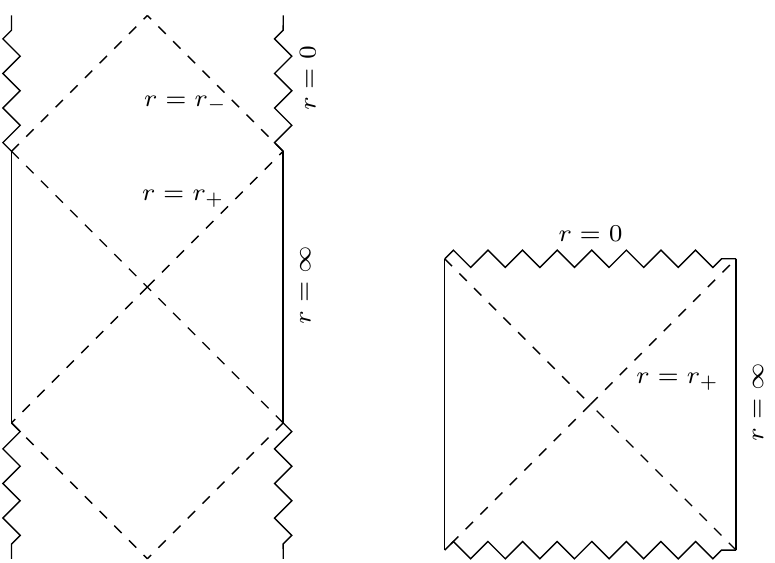}
\end{center}
\caption[Penrose diagrams of the BTZ.]{\label{fig:CPdiagrams} Penrose diagrams of the BTZ black hole for the rotating $0<r_-<r_+$ (left) and the static $0=r_-<r_+$ (right) cases.}
\end{figure}

%%%%%%%%%%%%%%%%

The BTZ black hole is a stationary, axisymmetric, (2+1) dimensional solution of the vacuum Einstein field equations with a negative cosmological constant $\Lambda=-1/\ell^2$ \cite{Banados:1992gq,Banados:1992wn}. It is diffeomorphic as a manifold to $M\equiv\bR\times I\times\mathbb{S}^1$, where $I$ is an open interval of the real line. Its metric $g$ can be realized in several, different, albeit equivalent ways, {\it e.g} by a suitable identification of points in the Poincar\'e patch of the three-dimensional AdS spacetime \cite{Banados:1992gq}. The ensuing line element reads
\begin{equation}\label{metric}
\dd s^2 = -N^2 \dd t^2 + N^{-2} \dd r^2 + r^2 \big(\dd\phi+ N^{\phi} \dd t \big)^2 \ ,
\end{equation}
where $t\in\mathbb{R}$, $\phi\in(0,2\pi)$, $r\in(r_+,\infty)$, while 
\begin{equation}\label{eq:metric_functions}
N^2 = -M+\frac{r^2}{\ell^2}+\frac{J^2}{4r^2} \ , \qquad N^\phi = -\frac{J}{2r^2} \ ,
\end{equation} 
$M$ being interpreted as the mass of the black hole and $J$ as its angular momentum.  The value of $r_+$ can be inferred, observing that, in the range $M>0$, $|J|\le M\ell$, $N$ vanishes at
\begin{equation}
r^2_{\pm}=
\frac{\ell^2}{2}
\left(
M \pm \sqrt{M^2-\frac{J^2}{\ell^2}}
\right) \ .
\end{equation}
These \textit{loci} are coordinate singularities and, thus, as customary in rotating black hole spacetimes, the BTZ solution possesses an inner $(r=r_-)$ and an outer horizon $(r=r_+)$. The Penrose diagrams of this spacetime are shown in Fig.~\ref{fig:CPdiagrams}. 

In addition, the event horizon turns out to be a Killing horizon whose generator reads
\begin{equation}\label{eq:Killing_Field}
\chi \doteq \partial_t +\Omega_{\mathcal{H}} \partial_\phi \ ,
\end{equation}
where $\Omega_{\mathcal{H}} \doteq N^\phi(r_{+}) = \frac{r_{-}}{\ell
r_{+}}$ is the angular velocity of the horizon. It is of paramount
relevance for this paper that $\chi$ is a
well-defined, {\em global, timelike Killing vector field} across the
whole exterior region $(r>r_+)$ of BTZ spacetime. This is the
sharpest difference in comparison to other models of rotating black hole
spacetimes, {\it e.g.}, the Kerr solution of the Einstein's equation
with vanishing cosmological constant. In these cases one is forced to
cope with the existence of a speed of light surface at which the
analogue Killing field is null.

In view of the distinguished role of $\chi$, it is natural to introduce the new coordinate system $(\tilde{t},r,\tilde{\phi})$, which is related to $(t,r,\phi)$ in such a way that $\partial_{\tilde{t}}=\chi$. The simplest choice consists of defining $t=\widetilde{t}$ and $\phi=\tilde{\phi}+\Omega_{\mathcal{H}}\tilde{t}$; the line element becomes
\begin{equation}\label{eq:new_line_element}
\dd s^2 = -N^2 \dd\tilde{t}^2 + N^{-2} \dd r^2 + \left(\dd\tilde{\phi}+(N^\phi+\Omega_{\mathcal{H}})\dd\tilde{t}\right)^2 \, .
\end{equation}
Observe that, while the range of $\tilde{t}$ is still $\bR$, that of $\tilde{\phi}$ is no longer simply the interval $(0,2\pi)$, rather $(-\Omega_\H \tilde{t}, 2\pi -\Omega_\H \tilde{t})$, with the end points still identified. Especially in the next section, we will be working mainly with \eqref{metric}, although, when we will be addressing the construction of a ground state, \eqref{eq:new_line_element} will turn out to be extremely useful.

% MASSIVE SCALAR FIELD IN BTZ

\section{Massive scalar field in BTZ}
\label{sec:KGeq}

\subsection{Klein-Gordon equation}
	
We consider a real, massive scalar field $\Phi: M\to\bR$ satisfying the Klein-Gordon equation,
	\begin{equation}\label{KG}
	P\Phi = (\Box_g - m^2-\xi R)\Phi =0\ ,
	\end{equation}
	where $\Box_g$ and $R$ are respectively the D'Alembert wave operator and the scalar curvature built out of \eqref{metric}, $\xi\in\bR$ while $m^2$ is the mass parameter of the scalar field. Since $R=-6/\ell^2$, it is convenient to introduce the dimensionless parameter $\mu^2 \doteq m^2 \ell^2-6\xi$. In addition, we assume that $m^2$ and $\xi$ are such that the Breitenlohner-Freedman bound $\mu^2 \geqslant -1$ holds \cite{Breitenlohner:1982jf}. 
	
	For our ultimate goal of quantizing \eqref{KG} and constructing the associated ground state(s) the first step in this direction consists of a careful study of the solutions of the Klein-Gordon equation. 
	Since the underlying spacetime is not globally hyperbolic, these cannot be constructed only by assigning initial data, for example on a constant time-$t$ hypersurface. One needs to supplement such information with the choice of an admissible boundary condition. A priori it is not obvious how to proceed since one might wish to assign such a condition either at the horizon $r=r_+$, at infinity $r\to\infty$ or possibly at both ends. This quandary is easily solved by showing that \eqref{KG} can be reduced to a second order ODE, whose boundary conditions are much easier to analyze. 
	
	To this end, we work with the coordinates $(t,r,\phi)$, so that \eqref{KG} reads
	\begin{gather} 
	 \left[-\frac{1}{N^2}\partial^2_t+\frac{1}{r}\partial_r\left(rN^2\right)\partial_r+\left(\frac{1}{r^2}-\frac{N^\phi}{N}\right)^2\partial^2_\phi \right. \notag\\
	+\left.2\frac{N^\phi}{N^2}\partial_t\partial_\phi-\frac{\mu^2}{\ell^2}\right]\Phi=0 \, . \label{eq:PDE_KG}
	\end{gather}
	
Since both $\partial_t$ and $\partial_\phi$ are Killing fields of \eqref{metric} we can take a Fourier expansion of $\Phi$,
	\begin{equation}
	\Phi(t,r,\phi)=\frac{1}{2\pi}\sum_{k\in\mathbb{Z}}\int_\mathbb{R} \dd\omega\;  e^{-i\omega t + i k \phi} \, \Psi_{\omega k}(r) \ .
	\end{equation}	
	It is convenient to introduce a new coordinate $z\in (0,1)$,
	\begin{equation}\label{eq:z_coordinate}
	z \doteq \frac{r^2-r_+^2}{r^2-r_-^2} \, ,
	\end{equation}
    so that, starting from \eqref{eq:PDE_KG}, $\Psi_{\omega k}(z)$ obeys
	\begin{equation} \label{eq:Sturm_Liouville_form}
		L_\omega \Psi_{\omega k}(z) \doteq \frac{\dd}{\dd z}\left(z\frac{\dd\Psi_{\omega k}(z)}{\dd z}\right)+q(z)\Psi_{\omega k}(z)=0 \, ,
	\end{equation}
	with
	\begin{multline}
		q(z) = \frac{1}{4(1-z)} \left[ \frac{\ell^2(\omega\ell r_{+} - k r_{-})^2}{(r_{+}^2-r_{-}^2)^2 z} \right. \\
		\left. -\frac{\ell^2(\omega\ell r_{-} - k r_{+})^2}{(r_{+}^2-r_{-}^2)^2}
		-\frac{\mu^2}{1-z} \right] \, ,
	\end{multline}
	This is indeed the sought second order ODE, written in Sturm-Liouville
	form, defined on the interval $(0,1)$. We need to clarify
	which are the admissible boundary conditions that can be assigned at
	$z=0$ (the horizon) or at $z=1$ (infinity). For ordinary differential
	equations this problem can be solved in full generality by using
	Sturm-Liouville theory, see {\it e.g.}, \cite{Zettl:2005,weidmann} or
	\cite{Dappiaggi:2016fwc} for an application to the study of a real,
	massive scalar field in the Poincar\'e patch of anti-de Sitter
	spacetime of arbitrary dimension. The nomenclature and the procedure
	that we will be using is the same employed in the last reference.
	For the sake of brevity, we will not recapitulate it fully here and we refer
	the reader to the works cited above.

\subsection{Solutions}
	
The next step consists of identifying a basis of the vector space of solutions of \eqref{eq:Sturm_Liouville_form}. Using Froebenius method, we infer that $\Psi_{\omega k}(z)=z^\alpha(1-z)^\beta F_{\omega k}(z)$, with 
\begin{equation}\label{eq:Frobenius_prequel}
\alpha^2=-\frac{\ell^4 r_+^2 \tilde{\omega}^2}{4 (r_{+}^2 - r_{-}^2)^2} \, , \quad \beta^2+\beta-\frac{\mu^2}{4}=0 \, ,
\end{equation}
where we define $\tilde{\omega} \doteq \omega-k\Omega_{\mathcal{H}}$ to be the square root of $\tilde{\omega}^2$ such that $\Im[\omega]=\Im[\tilde{\omega}] \geqslant 0$. By setting
	\begin{equation}\label{eq:Frobenius}
	\alpha = -i \, \frac{\ell^2 r_+ \tilde{\omega}}{2 (r_{+}^2 - r_{-}^2)}\ , \quad
	\beta = \frac{1}{2}\left(1 + \sqrt{1+\mu^2}\right)
	\end{equation}
	and plugging $\Psi_{\omega k}(z)$ in \eqref{eq:Sturm_Liouville_form}, we obtain the Gaussian hypergeometric equation,
	\begin{equation} \label{radialeq}
	z(1-z)\partial^2_z F_{\omega k} + [c -(a+b+1)z]\partial_z F_{\omega k} -ab F_{\omega k} =0 ,
	\end{equation}
	where
	\begin{equation}\label{eq:parameters}
	\begin{cases}
	\displaystyle a= \frac{1}{2}\left(1 + \sqrt{1 + \mu^2} - i\ell \, \frac{\tilde{\omega}\ell}{r_{+}-r_{-}} + i\ell\frac{k}{r_+}\right)\ , \\
	\displaystyle b= \frac{1}{2}\left(1 + \sqrt{1 + \mu^2} - i\ell \, \frac{\tilde{\omega}\ell}{r_{+} + r_{-}} - i\ell\frac{k}{r_+}\right)\ , \\
	\displaystyle c= 1 - i \, \frac{\ell^2 r_+ \tilde{\omega}}{r_{+}^2 - r_{-}^2}\ .
	\end{cases}
	\end{equation}
	For future convenience, we note that under the substitution $\tomega \mapsto \overline{\tomega}$, these parameters behave as
	\begin{equation}\label{eq:parameters-cc}
	\begin{aligned}
		a &\mapsto \overline{b-c+1} , & \alpha &\mapsto -\overline{\alpha} , \\ 
		b &\mapsto \overline{a-c+1} , & \beta &\mapsto \beta , \\
		c &\mapsto \overline{2-c} .
	\end{aligned}
	\end{equation}
	
	Generic solutions of \eqref{radialeq} can be written in closed form in terms of Gaussian hypergeometric functions that depend on the three parameters $a,b$ and $c$ of the equation. When choosing two linearly independent solutions, the dependence on these parameters forces us to disentangle two cases, accordingly to the values of $\mu^2$.

	\subsubsection{General case: $\mu^2\neq (n-1)^2-1, \ n=1,2,3,\dots$}	
	
	In this case, we choose as basis of solutions 
	\begin{subequations}\label{eq:z1gen}
	\begin{align}
	\Psi_1(z) & =z^\alpha(1-z)^\beta F(a,b,a+b-c+1;1-z) \label{Psi1a}  \, , \\
	\Psi_2(z) & =z^\alpha(1-z)^{1-\beta}\notag \\
	&\quad \times F(c-a,c-b,c-a-b+1;1-z) \label{Psi2a} \, ,
	\end{align}
	\end{subequations}
	For future reference and inspired by the terminology used for Sturm-Liouville problems \cite{Zettl:2005}, we call $\Psi_1$  the \emph{principal solution} at $z=1$, that is, the unique solution (up to scalar multiples) such that $\lim_{z\to 1} \Psi_1(z)/\Psi(z)=0$ for every solution $\Psi$ that is not a scalar multiple of $\Psi_1$. Note that $\Psi_2$ is not defined when the third argument becomes a non positive integer \cite{NIST}. From \eqref{eq:parameters}, we get $\sqrt{1+\mu^2}+1\not\in\mathbb{N}\cup\{0\}$, which identifies exactly the special range of values for $\mu^2$ which has been excluded. Observe in particular that this set includes the case $\mu^2=-1$ which saturates the BF bound \cite{Breitenlohner:1982jf}. This is a very special case, which would require a lengthy analysis on its own. For this reason, we will not consider it further in this paper.	

	Note that the above definitions obey $\Psi_1 \mapsto
	\overline{\Psi_1}$ and $\Psi_2 \mapsto \overline{\Psi_2}$ under the
	substitution $\tomega \mapsto \overline{\tomega}$. This can be checked
	using the conjugation identities~\eqref{eq:parameters-cc}, the
	symmetry $F(a,b,c;z) = F(b,a,c;z)$ and the second
	equality from (15.10.13) of~\cite{NIST}:
	\begin{multline*}
		F\left(a,b, a+b-c+1;1-z\right) \\
		=z^{1-c}F\left(a-c+1,b-c+1, a+b-c+1;1-z\right) .
	\end{multline*}

	\subsubsection{Special cases: $\mu^2 = (n-1)^2-1, \ n=2,3,\dots$}
	
	In this case, we choose the following basis of solutions for \eqref{eq:Sturm_Liouville_form} (see \cite[\S 15.10.8]{NIST}):
	\begin{subequations} \label{eq:z1spec}
	\begin{align}
	\Psi_1(z) & =z^\alpha(1-z)^\beta F(a,b,n;1-z) \label{Psi1b} \, , \\
	\Psi_2(z) & =z^\alpha(1-z)^{\beta} \notag \\
	&\quad \times \left[ F(a,b,n;1-z) \log(1-z) + K_n(z) \right]\label{Psi2b} \, ,
	\end{align}
	\end{subequations}
	where
	\begin{align}
	K_n(z) &= -\sum_{p=1}^{n-1}\frac{(n-1)!(p-1)}{(n-p-1)!(1-a)_p(1-b)_p}(z-1)^{-p} \notag \\
		 &\quad +\sum_{p=0}^{\infty} \frac{(a)_p(b)_p}{(n)_p p!} f_{p,n} \, (1-z)^p \, , \label{eq:G}
	\end{align}
	while $(a)_p = \Gamma(a+p)/\Gamma(a)$,
	\begin{equation*}
	f_{p,n} = \psi(a+p)+\psi(b+p) -\psi(1+p)-\psi(n+p) \, ,
	\end{equation*}
	and $\psi$ is the digamma function. Observe that, also in these special cases, $\Psi_1$ is the principal solution at $z=1$. 

	Note that $\Psi_1 \mapsto \overline{\Psi_1}$ under the substitution $\tomega \mapsto \overline{\tomega}$, by the same argument as in the generic case. We do not need to check this property for $\Psi_2$ since, as it will be clear in next section, $\Psi_1$ is the only solution which plays a role for the admissible boundary conditions.

\subsection{End point classification} 
\label{sec:endpoints}

Having specified a basis of solutions of \eqref{eq:Sturm_Liouville_form}, we can continue in our quest to identify the admissible boundary conditions at the end points 0 and 1 for \eqref{KG}. These will depend on the square integrability of the solutions near the end points, in a completely analogous way to the case of the Poincar\'e patch of AdS analyzed in \cite{Dappiaggi:2016fwc}.

We start by identifying the fall-off behavior of the solutions of \eqref{eq:Sturm_Liouville_form} separately at the end points $z=0$ and at $z=1$. This allows to classify the end points in the following way: we call the end point 0 (respectively 1) \emph{limit circle} if, for some $\tilde{\omega} \in \bC$, all solutions of \eqref{eq:Sturm_Liouville_form} are in $L^2((0,z_0);\mathcal{J}(z)\dd z)$ for some $z_0 \in (0,1)$ [respectively $L^2((z_1,1);\mathcal{J}(z)\dd z)$ for some $z_1 \in (0,1)$]; otherwise, we call it \emph{limit point}. The measure $\mathcal{J}(z)\dd z$, with
\begin{equation} \label{eq:measure}
\mathcal{J}(z) = \frac{1}{1-z}+\frac{r^2_+}{z(r^2_+-r^2_-)} \, ,
\end{equation}
satisfies the relation $\dd\nu(g)=\pi_I^*(\mathcal{J}(z)\dd z) \dd \varphi$, where $\dd \nu(g) = r/N^2 \, \dd r \dd\varphi$ and $\pi_I:M\to I$ is the projection along the $z$-direction. Notice that the operator $S_\tomega\Psi(z) \doteq \frac{1}{\mathcal{J}(z)} L_\tomega\Psi(z)$, with $L_\tomega$ from \eqref{eq:Sturm_Liouville_form}, is Hermitian with respect to the measure $\mathcal{J}(z)\dd z$.

A direct inspection of \eqref{Psi1a} and \eqref{Psi2a} as well as of \eqref{Psi1b} and \eqref{Psi2b}, supplemented with the asymptotic behavior of the hypergeometric function at $z=0$ and $z=1$, yields the sought result for the basis elements of the space of solutions of \eqref{eq:Sturm_Liouville_form}. For convenience we summarize the results described below in table \ref{tab:BCsummary}.

\subsubsection{End point $z=1$}

	At $z=1$, since the hypergeometric function is equal to $1$ when evaluated at the origin, the behavior of \eqref{Psi1a} and \eqref{Psi2a} can be inferred from that of $(1-z)^\beta$ and $(1-z)^{1-\beta}$ respectively. By accounting also for the integration measure and using \eqref{eq:Frobenius}, it turns out that $\Psi_1$ lies in $L^2((z_1,1);\mathcal{J}(z)\dd z)$ for all values of $\mu^2>-1$ and regardless of $z_1\in(0,1)$ and of $\tilde{\omega}$. On the contrary, $\Psi_2$ lies in $L^2((z_1,1);\mathcal{J}(z)\dd z)$ if $-1<\mu^2<0$, again regardless of $z_1\in(0,1)$ and of $\tilde{\omega}$. Therefore, we say that $z=1$ is limit point if $\mu^2\geqslant 0$ while it is limit circle if $-1<\mu^2<0$.
	
	For the special cases $\mu^2=(n-1)^2-1$, $n=2,3,\ldots$, the first basis element $\Psi_1$ as in \eqref{Psi1b} behaves exactly like \eqref{Psi1a}. At the same time, $\Psi_2$ as in \eqref{Psi2b} never lies in $L^2((z_1,1);\mathcal{J}(z)\dd z)$ on account of the singularities of $K_n(z)$. Hence, $z=1$ is always limit point of $\mu^2 \geqslant 0$.

\subsubsection{End point $z=0$}
\label{sec:endpoint0}

	In order to understand the behavior of the solutions of \eqref{eq:Sturm_Liouville_form} at $z=0$, we need to consider a different, more convenient basis,
	\begin{subequations} \label{Psi0}
	\begin{align}
	\Psi_3(z) &= z^\alpha(1-z)^\beta F(a,b,c;z) \, , \label{Psi10}  \\
	\Psi_4(z) &= z^{-\alpha}(1-z)^\beta \notag \\
	&\quad \times F(a-c+1,b-c+1,2-c;z)  \label{Psi20} \, .
	\end{align}
	\end{subequations}
	where $a,b,c$ are defined in \eqref{eq:parameters}. Observe that $\Psi_3$ and $\Psi_4$ form a well-defined basis of solutions for all $\mu^2 > -1$, except when $c=1$ ($\alpha = 0$), whose case is dealt with separately below.
	
	Since the hypergeometric function is equal to $1$ when evaluated at $z=0$, the leading behavior of the two solutions at the origin is regulated by $z^\alpha$ in the first case and by $z^{-\alpha}$ in the second one. It is easy to verify that $\Psi_3 \in L^2((0,z_0),\mathcal{J}(z)\dd z)$ for $\Im[\tilde{\omega}]>0$, irrespectively of $z_0\in(0,1)$, while $\Psi_4 \in L^2((0,z_0),\mathcal{J}(z)\dd z)$ if $\Im[\tilde{\omega}]<0$. For $\Im[\tilde{\omega}]=0$ none of the solutions is square integrable since a logarithmic singularity occurs. Therefore, we say that $z=0$ is limit point.  

If $c=1$, then $\omega=k\frac{r_-}{\ell r_+} = k \Omega_{\mathcal{H}}$ satisfies a synchronization condition with the black hole angular velocity, a case extensively studied in \cite{Ferreira:2017}. The solutions $\Psi_3$ and $\Psi_4$ no longer form a basis of solutions of \eqref{eq:Sturm_Liouville_form}, hence, we consider the following basis \cite[\S 15.10.8]{NIST}:
	\begin{subequations}
	\begin{gather*}
	(1-z)^\beta F(a,b,1;z) \, ,  \label{Psi01b} \\
	(1-z)^\beta\left[F(a,b,1;z)\log(z)+K_1(1-z)\right] \, , \label{Psi02b}
	\end{gather*}
	\end{subequations}
where $K_1$ is as in \eqref{eq:G}. A close inspection of these two solutions unveils that the leading behavior at $z=0$ is dominated by a constant in the first case and by $\log (z)$ in the second one. Hence, none of the solutions lies in $L^2((0,z_0),\mathcal{J}(z)\dd z)$ regardless of $z_0\in(0,1)$. This is in agreement with the previous point.

%
%%% TABLE
%
\setlength{\tabcolsep}{2ex}           % space between columns
\renewcommand{\arraystretch}{1.25}    % space between rows
\begin{center}
	\begin{table*}[t]
		\begin{tabular}{cccc}
		\hline\hline
		Range of $\mu^2$ & Range of $\tilde{\omega}$ & $L^2$ at $z=0$ & $L^2$ at $z=1$ \\
		\hline
		& $\Im[\tilde{\omega}] > 0$ & $\Psi_3$ & $\Psi_1$ and $\Psi_2$ \\
		$-1<\mu^2<0$ & $\Im[\tilde{\omega}] = 0$ & none & $\Psi_1$ and $\Psi_2$ \\
		& $\Im[\tilde{\omega}] < 0$ & $\Psi_4$ & $\Psi_1$ and $\Psi_2$ \\
		\hline
		& $\Im[\tilde{\omega}] > 0$ & $\Psi_3$ & $\Psi_1$ \\
		$\mu^2 \geqslant 0$ & $\Im[\tilde{\omega}] = 0$ & none & $\Psi_1$ \\
		& $\Im[\tilde{\omega}] < 0$ & $\Psi_4$ & $\Psi_1$ \\
		\hline\hline
		\end{tabular}
		\caption{Summary of the square integrability at $z=0$ and at $z=1$ of a basis of solutions for \eqref{eq:Sturm_Liouville_form} depending on the parameters $\mu^2$ and $\tilde{\omega}$ of the equation. The integration measure is $\mathcal{J}(z)\dd z$ as per \eqref{eq:measure}. \label{tab:BCsummary}}
	\end{table*}
\end{center}

\vspace*{-5.95ex}
Note that for $\tomega \not\in \bR$, hence excluding the $c=1$ case, the above definitions obey $\Psi_3 \mapsto
\overline{\Psi_4}$ and $\Psi_4 \mapsto \overline{\Psi_3}$ under the substitution $\tomega \mapsto \overline{\tomega}$. This can be checked using the conjugation identities~\eqref{eq:parameters-cc} and the symmetry $F(a,b,c;z) = F(b,a,c;z)$.

\subsection{Robin boundary conditions}
		
We can address finally the question of which are the admissible boundary conditions at the two end points $z=0$ and $z=1$. Tentatively, as in the simple example of a massive scalar field in the Poincar\'e patch of AdS studied in \cite{Dappiaggi:2016fwc}, we wish to impose Robin boundary conditions at $z=1$ for a range of the mass parameter $\mu^2$ of the scalar field. In fact, as pointed out in \cite{Ferreira:2017}, imposing Robin boundary conditions is equivalent to requiring zero energy flux through the conformal boundary, a natural physical condition.

To start with we focus our attention on the ODE \eqref{eq:Sturm_Liouville_form} at fixed value of $\tomega$ and $k$. Since we deal with a singular Sturm-Liouville problem, it is not possible to assign Robin boundary conditions by specifying the value of a linear combination between a solution and its derivative. This statement is supported also by the observation that at $z\to 1$ both solutions $\Psi_2(z)$ as per \eqref{Psi2a} and per \eqref{Psi2b} are divergent.
	
This problem can be overcome by using Sturm-Liouville theory. While we do not wish to enter in a full explanation of the technical details, which are fully accounted for in \cite{Dappiaggi:2016fwc} and in \cite{Zettl:2005}, we outline the main idea of the procedure. The rationale consists of observing that, in a so-called regular Sturm-Liouville problem, a generic Robin boundary condition can be expressed equivalently either in terms of a linear combination between a solution and its derivative or in terms of a linear combination between the Wronskians of such solution with respect to two linearly independent solutions, one of which is chosen to be the principal solution.
	
In the case at hand, this translates to the following: we say that a solution $\Psi_{\zeta}$ of \eqref{eq:Sturm_Liouville_form} satisfies a {\em Robin boundary condition} at $z=1$ parametrized by $\zeta \in [0,\pi)$ if
\begin{equation} \label{eq:RBC}
\lim_{z \to 1} \left\{ \cos(\zeta) \mathcal{W}_z[\Psi_{\zeta}, \Psi_1] + \sin(\zeta) \mathcal{W}_z[\Psi_{\zeta}, \Psi_2] \right\} = 0 \, ,
\end{equation}
where $\Psi_1$ is the principal solution at $z=1$ [\eqref{Psi1a} or \eqref{Psi1b}], $\Psi_2$ is a second linearly independent solution [for instance, \eqref{Psi2a} or \eqref{Psi2b}] and both are square integrable in a neighborhood of $z=1$. Here, $\mathcal{W}_z[u,v] \doteq u(z)v'(z) - v(z)u'(z)$ is the Wronskian computed with respect to two differentiable functions $u$ and $v$. As a consequence, the solution $\Psi_{\zeta}$ may be written as
\begin{equation}\label{eq:Robin_solution}
\Psi_{\zeta}(z) = \cos(\zeta)\Psi_1(z)+\sin(\zeta)\Psi_2(z) \, .
\end{equation}

We note that $\zeta=0$ corresponds to the standard Dirichlet boundary condition since it guarantees that $\Psi_{\zeta}$ coincides with $\Psi_1$. At the same time, if $\zeta=\frac{\pi}{2}$, we say that $\Psi_{\zeta}$ satisfies a Neumann boundary condition, coinciding with $\Psi_2$. Yet, contrary to the Dirichlet boundary condition, this is not a universal assignment as it depends on the choice of $\Psi_2$. 

The requirement of square-integrability of both $\Psi_1$ and $\Psi_2$ near $z=1$ implies that a Robin boundary condition can only be applied when $-1 < \mu^2 < 0$, as analyzed in the last section. For $\mu^2 \geqslant 0$, only the principal solution $\Psi_1$ is square integrable near $z=1$ and, hence, no boundary condition is required. In practice, this is as if the Dirichlet boundary condition had been chosen.

A similar reasoning could be applied at $z=0$, but, as we have shown in the preceding subsection, if we focus only on square integrable solutions, only one exists, provided that $\Im[\tilde{\omega}] \neq 0$. Therefore, at $z=0$ there is no need to impose any boundary condition.

% TWO-POINT FUNCTION

\section{Two-point Function}
\label{Sec:ground_state}

In this section, we address the main question of this paper, namely the construction of a class of two-point functions, investigating whether they define a ground state for a real, massive scalar field in the BTZ black hole spacetime. We will follow the same procedure employed in \cite{Dappiaggi:2016fwc} in the Poincar\'e patch of an AdS spacetime of arbitrary dimension. As we will point out in the subsequent discussion, the main structural difference lies in the underlying metric being stationary, unless one considers the static case ($J=0$) in \eqref{eq:metric_functions}. 

Dropping for the moment the requirement of individuating a ground state, in general, by {\em two-point function} (or Wightman function) we refer to a bidistribution $G^+\in\mathcal{D}^\prime(M\times M)$ such that
\begin{equation}\label{eq:eom_G}
(P\otimes\mathbb{I})G^+=(\mathbb{I}\otimes P)G^+=0 \, ,
\end{equation}
and 
\begin{equation}\label{eq:pos_G}
G^+(f,f)\geq 0 \, , \quad \forall f\in C^\infty_0(M) \, .
\end{equation}
In addition, the antisymmetric part of $G^+$ is constrained to coincide with the commutator distribution, in order to account for the canonical commutation relations (CCRs) of the underlying quantum field theory.

In order to make this last requirement explicit, let us consider the coordinate system $(t,z,\phi)$ introduced in \eqref{metric} with $r$ replaced by $z$ as in \eqref{eq:z_coordinate}. Working at the level of the integral kernel for $G^+$ and imposing the CCRs is tantamount to requiring that the antisymmetric part  $iG(x,x^\prime)$, $x,x^\prime\in M$, where 
$$ i G(x,x^\prime) = G^+(x,x^\prime)-G^+(x^\prime,x)$$
satisfies \eqref{eq:eom_G} together with the initial conditions
\begin{subequations} \label{eq:initial_conditions_E}
\begin{align} 
G(x,x^\prime)|_{t=t^\prime} &= 0, \label{eq:initial_conditions_E_1} \\
-\partial_t  G(x,x^\prime)|_{t=t^\prime} &=\partial_{t^\prime}G(x,x^\prime)|_{t=t^\prime}=\frac{\delta(z-z^\prime)\delta(\phi-\phi^\prime)}{\mathcal{J}(z)}, \label{eq:initial_conditions_E_2}
\end{align}
\end{subequations}
with $\mathcal{J}(z)$ as in \eqref{eq:measure}.

In order to construct explicitly the two-point function we assume that $G^+$ admits a mode expansion
\begin{align}
G^+(x,x^\prime) &= \lim_{\epsilon \to 0^+} \sum\limits_{k\in\mathbb{Z}} \int_\bR \frac{\dd\omega}{(2\pi)^2} \, e^{-i\omega (t-t^\prime-i\epsilon)+ik(\phi-\phi^\prime)} \notag \\
&\quad \times \widehat{G}_{\omega k}(z,z^\prime) \, , \label{eq:2-pt_modes}
\end{align}
where $x,x^\prime\in M$, $i\epsilon$ has been added as a regularization while the limit has to be taken in the weak sense. At this point, it is convenient to recall that, although both $\partial_t$ and $\partial_\phi$ are global Killing vector fields, a more prominent physical role is played by the globally timelike Killing vector field $\chi$ defined in \eqref{eq:Killing_Field}. More precisely, in the construction of a ground state, the notion of positive frequencies is played by $\tilde{\omega}=\omega-k\Omega_{\mathcal{H}}$ which is subordinated to $\chi$. Hence, in order to make the role of $\tilde{\omega}$ manifest, following the discussion of Section \ref{sec:BTZ_geometry}, we change from $(\omega, k)$ to $(\tilde{\omega},k)$ and from the coordinates $(t,r,\phi)$ to $(\tilde{t},r,\tilde{\phi})$, where $\tilde{\phi}=\phi-\Omega_{\mathcal{H}} t$ and $\tilde{t}=t$. Moreover, since only the positive $\tilde{\omega}$-frequencies contribute to the two-point function of the ground state, we can write $\widehat{G}_{\omega k}(z,z^\prime) \doteq \widetilde{G}_{\tilde{\omega} k}(z,z^\prime) \Theta(\tilde{\omega})$, with $\widetilde{G}_{\tilde{\omega} k}(z,z^\prime)$ defined for all $\tilde{\omega} \in \bR$.

Taking into account these comments and recalling that the antisymmetric part ought to satisfy \eqref{eq:initial_conditions_E_1}, a natural requirement consists of looking for $\widetilde{G}_{\tilde{\omega} k}(z,z^\prime)$ which is symmetric for exchange of $z$ and $z^\prime$ and such that $\widetilde{G}_{-\tilde{\omega},-k}(z,z^\prime) = -\widetilde{G}_{\tilde{\omega} k}(z,z^\prime)$. In this way, the commutator distribution reads
\begin{align}
i G(x,x^\prime) &= \lim_{\epsilon \to 0^+} \sum\limits_{k\in\mathbb{Z}} \int_\bR \frac{\dd\tilde{\omega}}{(2\pi)^2} e^{-i\tilde{\omega} (t-t^\prime-i|\tomega|\epsilon) + ik(\tilde{\phi}-\tilde{\phi^\prime})} \notag \\
&\quad \times  \widetilde{G}_{\tilde{\omega} k}(z,z^\prime) \, , \label{eq:propagator_modes}
\end{align}
where $\widetilde{G}_{\omega k}(z,z^\prime)$ is a mode bidistribution chosen in such a way that, \textit{c.f.}~Eq.~\eqref{eq:initial_conditions_E_2},
\begin{equation} \label{eq:delta-resolution}
\int_\bR \frac{\dd\tilde{\omega}}{2\pi} \, \tilde{\omega} \, \widetilde{G}_{\tilde{\omega} k}(z,z^\prime) = \frac{\delta(z-z^\prime)}{\mathcal{J}(z)} \, .
\end{equation}
This identity, together with the Fourier series for the delta distribution along the angular coordinates, guarantees that finding $\widetilde{G}_{\tilde{\omega} k}(z,z^\prime)$ is tantamount to constructing a full-fledged two-point function $G^+$, provided that positivity as in \eqref{eq:pos_G} is satisfied. In addition, \eqref{eq:eom_G} entails that the mode bidistribution is such that
$$(L_\tomega\otimes\mathbb{I})\widetilde{G}_{\tilde{\omega} k}(z,z^\prime)=(\mathbb{I}\otimes L_\tomega)\widetilde{G}_{\tilde{\omega} k}(z,z^\prime)=0 \, ,$$
where $L_\tomega$ is defined in \eqref{eq:Sturm_Liouville_form}. 

Our next goal will be to use this information to construct explicitly $\widetilde{G}_{\tomega k}(z,z^\prime)$ in terms of solutions of \eqref{eq:Sturm_Liouville_form}. Our strategy, as in \cite{Dappiaggi:2016fwc}, will be to obtain an integral representation for the delta distribution on the RHS of \eqref{eq:delta-resolution}, from which we can read off $\widetilde{G}_{\tilde{\omega} k}(z,z^\prime)$. However, and contrarily to the case of pure AdS analyzed in \cite{Dappiaggi:2016fwc}, we face a technical hurdle. When dealing with the static case $J=0$, the ODE \eqref{eq:Sturm_Liouville_form} can be treated as an eigenvalue problem with spectral parameter $\tilde{\omega}^2$ and it is possible to express the delta distribution as an expansion in terms of the eigenfunctions of $L_\tomega$ (resolution of the identity). But this is not possible when dealing with the non static case $J \neq 0$, in which case the ODE \eqref{eq:Sturm_Liouville_form} has linear terms in $\tilde{\omega}$. Instead, we may treat $L_\tomega$ as a \emph{quadratic operator pencil}, \emph{i.e.}~a differential operator with quadratic dependence on the spectral parameter $\tilde{\omega}$. In Appendix~\ref{apx:deltaexpansion}, it is described how to obtain the expansion of the delta distribution in terms of eigenfunctions of an operator of this type.

In the following, we present the results for the resolution of the identity and for the mode expansion of the two-point function for a fixed Robin boundary condition.
We start from the simplest scenario, $\mu^2\geqslant 0$, for which no boundary condition needs to be imposed to the solutions of \eqref{eq:Sturm_Liouville_form} at $z=1$, and then consider the more interesting case $-1 < \mu^2 < 0$. The full details of the calculation can be consulted in Appendix~\ref{apx:calculation-delta-expansion}.

\subsection{Case $\mu^2\geqslant 0$}

For $\mu^2 \geqslant 0$ both $z=0$ and $z=1$ in the Sturm-Liouville problem associated to \eqref{eq:Sturm_Liouville_form} are of limit point type. Using the results of Appendix~\ref{apx:calculation-delta-expansion} in the case $\zeta=0$, it is possible to obtain an integral representation of $\delta(z-z')$ in terms of eigenfunctions of $L_\tomega$,
$$\frac{\delta(z-z^\prime)}{\mathcal{J}(z)} = \int_{\bR} \frac{\dd\tilde{\omega}}{2\pi i} \, \tilde{\omega}  \left(\frac{A}{B}-\frac{\overline{A}}{\overline{B}}\right) C \, \Psi_1(z)\Psi_1(z^\prime) \, , $$
where the constants $A$, $B$ and $C$ are defined as
\begin{subequations} \label{eq:A_B_constants}
\begin{align}
A &= \frac{\Gamma(c-1)\Gamma(c-a-b)}{\Gamma(c-a)\Gamma(c-b)} \, , \\
B &=\frac{\Gamma(c-1)\Gamma(a+b-c)}{\Gamma(a)\Gamma(b)} \, , \\
C &= \frac{\ell^4}{4(r_+^2-r_-^2)\sqrt{1+\mu^2}} \, .
\end{align}
\end{subequations}
Comparing with \eqref{eq:delta-resolution}, we can read off $\widetilde{G}_{\tilde{\omega} k}(z,z^\prime)$ and write the two-point function as
%
%\begin{widetext}
\begin{align} 
G^+(x,x^\prime) &= \lim_{\epsilon \to 0^+} \sum\limits_{k\in\mathbb{Z}} e^{ik\left(\tilde{\phi}-\tilde{\phi}^\prime\right)}\int_0^\infty \frac{\dd\tilde{\omega}}{(2\pi)^2} \, e^{-i\tilde{\omega}\left(\tilde{t}-\tilde{t}^\prime-i\epsilon\right)} \notag
\\
&\quad \times \left(\frac{A}{B}-\frac{\overline{A}}{\overline{B}}\right) C \, \Psi_1(z)\Psi_1(z^\prime) \, . \label{eq:state_Dirichlet_v2}
\end{align}
%\end{widetext}
%
The mode decomposition of $G^+$ in \eqref{eq:state_Dirichlet_v2} contains only positive $\tilde{\omega}$-frequencies and, per construction, its antisymmetric part satisfies \eqref{eq:initial_conditions_E}. Hence, it is legitimate to call the state associated with $G^+$ the {\em ground state} for a real, massive scalar field in the BTZ spacetime with $\mu^2 \geqslant 0$. 

An important related question consists of whether $G^+$ is
locally of Hadamard form. Such property is desirable not only at a
structural level but also for constructing Wick polynomials, the
building blocks for dealing with interactions at a perturbative level.
In Ref.~\cite{Sahlmann:2000fh} it is proven under rather general
hypotheses that a ground state, such as the one defined by
\eqref{eq:state_Dirichlet_v2} in particular, is always of \emph{local}
Hadamard form, namely $G^+$ identifies a Hadamard state in every
globally hyperbolic subregion of BTZ (for the definition of Hadamard
state refer to Ref.~\cite{Khavkine:2014mta}). A more difficult task is
to verify if this ground state satisfy a \emph{global} Hadamard
condition such as the one proposed in \cite{Dappiaggi:2016fwc} and
\cite{Dappiaggi:2017wvj} for a quantum state in anti-de Sitter
spacetime. Although we conjecture that to be the case, we leave a
rigorous verification for future work.

\subsection{Case $-1<\mu^2<0$}
	
For $-1<\mu^2<0$, a Robin boundary condition needs to be imposed on solutions at $z=1$ and therefore the analysis of the previous section is changed as we obtain a different two-point function for each possible Robin boundary condition. We have to consider separately two regimes, $\zeta\in[0,\zeta_*)$ and $\zeta\in[\zeta_*,\pi)$, with
\begin{align} \label{eq:zetacritical}
\zeta_* \doteq \arctan\left( \frac{\Gamma\left(2\beta-1\right)\left|\Gamma\left(1-\beta+i\ell\frac{k}{r_+}\right)\right|^2}{\Gamma\left(1-2\beta\right)\left|\Gamma\left(\beta+i\ell\frac{k}{r_+}\right)\right|^2} \right) \, ,
\end{align}
where $\beta = \frac{1}{2}+\frac{1}{2}\sqrt{1+\mu^2}$ was defined in \eqref{eq:Frobenius}. Since $\mu^2 \in (-1,0)$ and thus $\beta \in (\frac{1}{2},1)$, it follows that $\zeta_* \in (\frac{\pi}{2}, \pi)$.

\subsubsection{Case $\zeta\in[0,\zeta_*)$}

For Robin boundary conditions such that $\zeta\in[0,\zeta_*)$, it turns out that the spectrum of the operator $L_\tomega$ in \eqref{eq:Sturm_Liouville_form} is only $\tilde{\omega} \in \bR$ and does not include any isolated eigenvalue in $\bC \setminus \bR$, which would correspond to poles in the Green's distribution associated with $L_\tomega$ (see Appendices~\ref{apx:calculation-delta-expansion} and \ref{apx:S2} for more details). Observe that, since $\zeta_* \in (\frac{\pi}{2}, \pi)$, this scenario includes both the Dirichlet and the Neumann boundary conditions. This situation is structurally identical to the one investigated in the previous section for $\mu^2 \geqslant 0$. Using the results of Appendix~\ref{apx:calculation-delta-expansion} we obtain the following resolution of the identity
\begin{equation} \label{eq:identity_resolution}
\frac{\delta(z-z^\prime)}{\cJ(z)} = \int_{\bR} \frac{\dd\tilde{\omega}}{2\pi i} \, \tilde{\omega} \, \frac{\left(A\overline{B}-\overline{A}B\right) C}{|{\cos(\zeta) B-\sin(\zeta)A}|^2} \Psi_{\zeta}(z)\Psi_{\zeta}(z') \, ,
\end{equation}
where the constants $A$, $B$ and $C$ are the same as in \eqref{eq:A_B_constants}. We can use this result in combination with \eqref{eq:2-pt_modes} and \eqref{eq:delta-resolution} to obtain, for each $\zeta\in [0,\zeta_*)$,
%
%\begin{widetext}
	\begin{align}
	G^+_\zeta(x,x^\prime) &= \lim_{\epsilon \to 0^+} \sum_{k\in\mathbb{Z}} e^{ik\left(\tilde{\phi}-\tilde{\phi}^\prime\right)} \int_0^{\infty} \frac{\dd\tilde{\omega}}{(2\pi)^2} \, e^{-i\tilde{\omega} \left(\tilde{t}-\tilde{t}^\prime-i\epsilon\right)} \notag 
	\\
	&\quad \times \frac{\left(A\overline{B}-\overline{A}B\right) C}{|{\cos(\zeta) B-\sin(\zeta)A}|^2} \Psi_{\zeta}(z)\Psi_{\zeta}(z') \, . \label{eq:state_Robin_no_BS}
	\end{align}
%\end{widetext}
%
Note that this two-point function, valid for scalar fields with $-1<\mu^2<0$, coincides with the one for scalar fields with $\mu^2 \geqslant 0$ obtained in \eqref{eq:state_Dirichlet_v2} if $\zeta=0$, that is, for Dirichlet boundary conditions.

%%% FIGURE
%
% Using figure* makes the figure be widetext-length

\begin{figure*}[t!]
	\centering
	\includegraphics[width=0.44\linewidth]{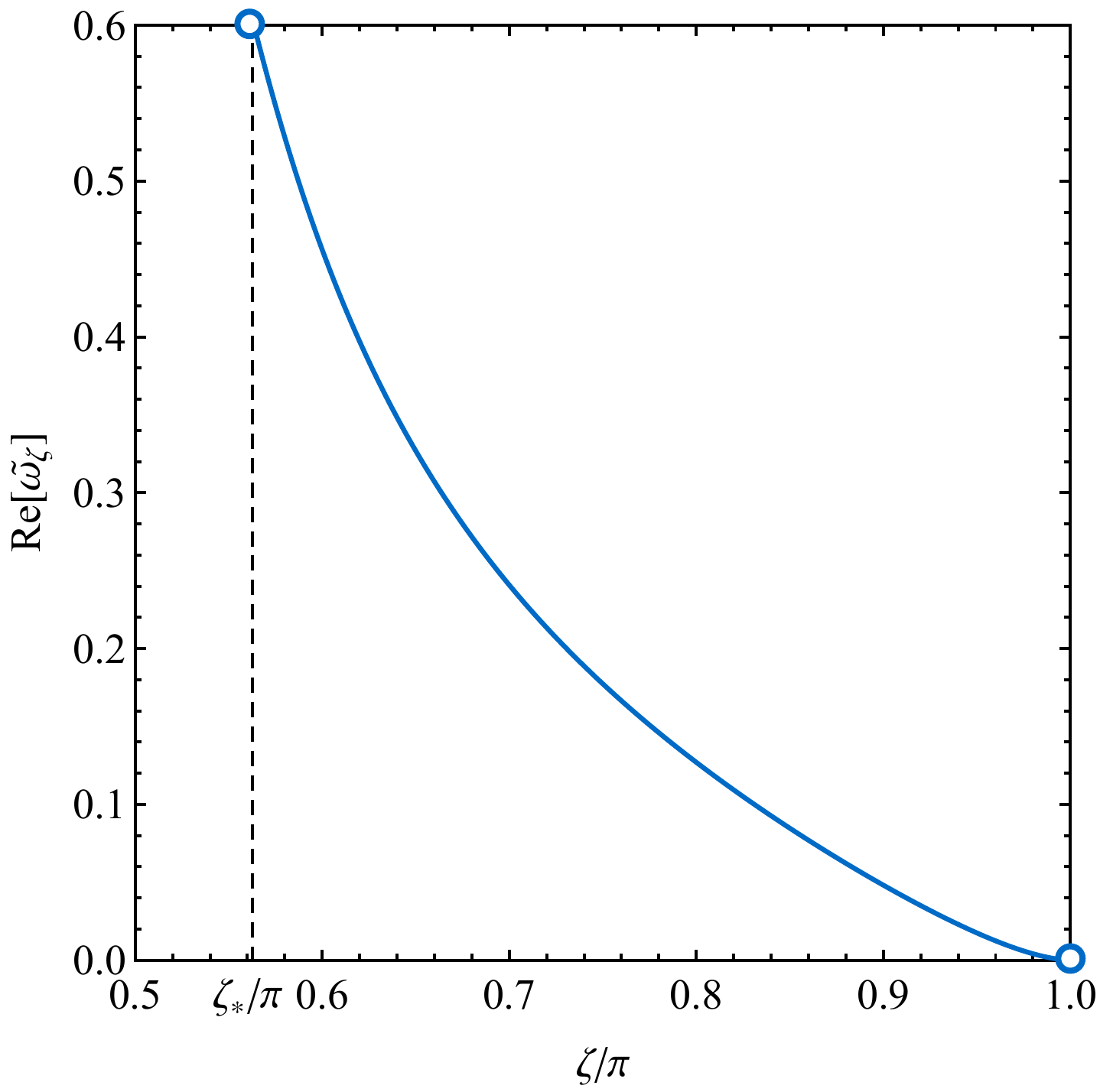} \hspace{8ex}
	\includegraphics[width=0.44\linewidth]{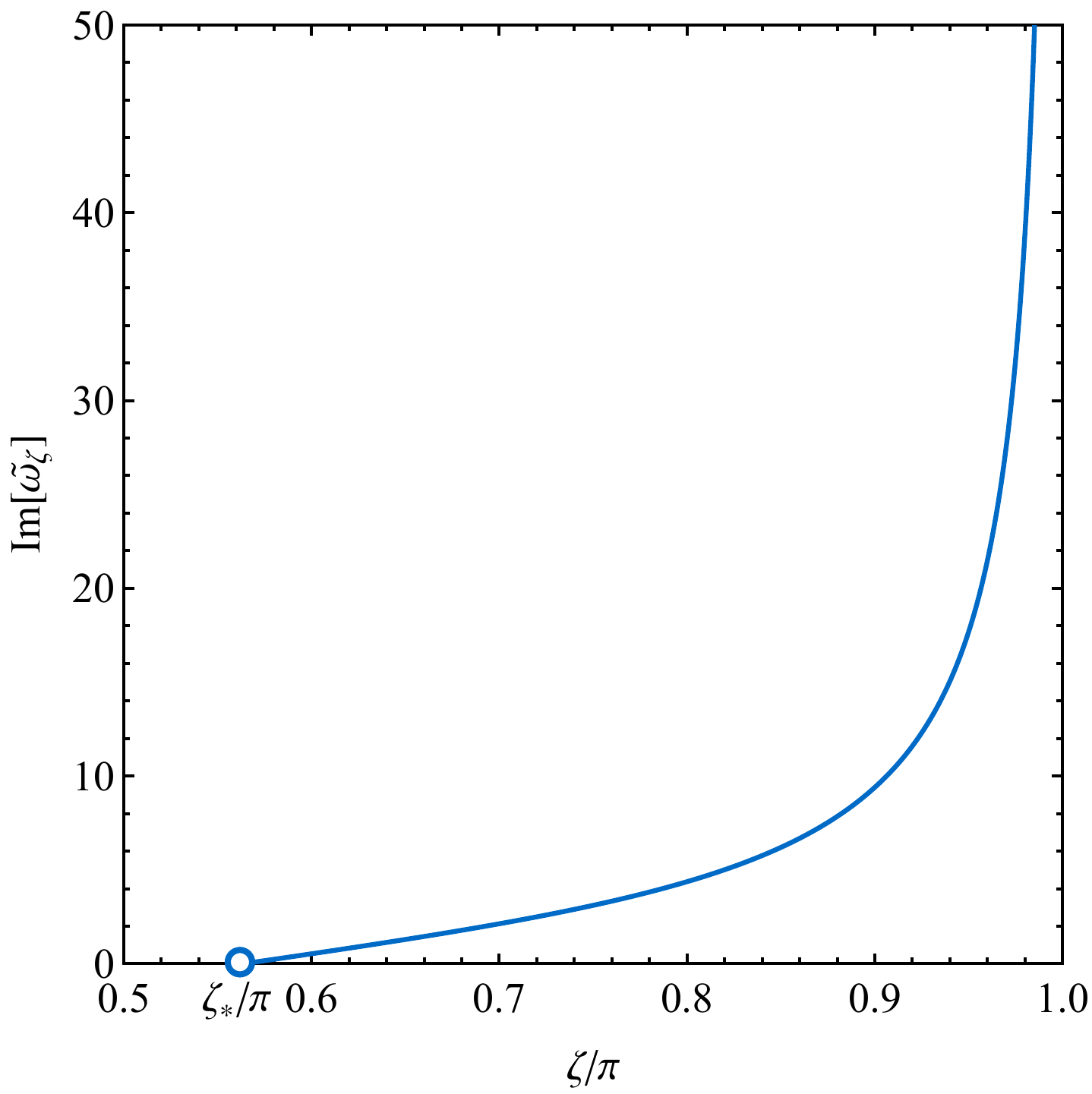}
	\caption{\label{fig:bound-state}
	Real and imaginary part of the bound state frequency $\tilde{\omega}_{\zeta}$ as a function of the parameter $\zeta$ defining the Robin boundary condition for a BTZ black hole with $\ell=1$, $r_+=5$ and $r_-=3$ and a scalar field with $\mu^2 = - 0.65$ and $k=1$. The bound state mode solutions exist for values of $\zeta$ between $\zeta_* \approx 0.5625\pi$ and $\pi$.}
\end{figure*}

\subsubsection{Case $\zeta\in[\zeta_*,\pi)$}

For Robin boundary conditions such that $\zeta\in[\zeta_*,\pi)$, it turns out that the spectrum of the operator $L_\tomega$ in \eqref{eq:Sturm_Liouville_form} not only contains all $\tilde{\omega} \in \bR$ but it includes also two isolated eigenvalues in $\bC \setminus \bR$, complex conjugate to each other, which correspond to poles in the Green's distribution associated with $L_\tomega$ (see Appendices~\ref{apx:calculation-delta-expansion} and \ref{apx:S2} for more details). Denote those eigenvalues by $\tilde{\omega}_\zeta$ and $\overline{\tilde{\omega}_\zeta}$ such that $\Im[\tilde{\omega}_\zeta]>0$. They are dubbed {\em bound state frequencies} and their corresponding eigensolutions are called {\em bound state mode solutions}. The existence of bound state mode solutions was also verified in \cite{Dappiaggi:2016fwc} for the case of a massive scalar field in the Poincar\'e patch of AdS when Robin boundary conditions parametrized with $\zeta \in (\frac{\pi}{2},\pi)$ are imposed at conformal infinity.

Unfortunately, an analytic expression for $\tilde{\omega}_\zeta$ cannot be found since, for $\Im[\tilde{\omega}_\zeta]>0$ and fixed $\zeta$, one needs to invert the equality
$$\tan(\zeta) = \left.\frac{B}{A}\right|_{\tilde{\omega} = \tilde{\omega}_\zeta} \, , $$
where the constants $A$ and $B$ are the same as in \eqref{eq:A_B_constants}. This operation can only be completed numerically (except in very particular cases such as $\zeta=0$ and $\zeta=\pi/2$) and a representative example is shown in Fig.~\ref{fig:bound-state}. A more qualitative discussion of the behavior of the solutions $\tomega_\zeta$ as a function of $\zeta$ can be found in Appendix~\ref{apx:S2}.

As a consequence of these bound state frequencies, the resolution of the identity acquires an extra term in comparison to \eqref{eq:identity_resolution}, which, following Appendix~\ref{apx:calculation-delta-expansion}, can be computed via Cauchy's residue theorem, yielding
\begin{align}
\frac{\delta(z-z^\prime)}{\mathcal{J}(z)} &= \int_{\bR} \frac{\dd\tilde{\omega}}{2\pi i} \, \tilde{\omega} \, \frac{\left(A\overline{B}-\overline{A}B\right) C}{|{\cos(\zeta) B-\sin(\zeta)A}|^2} \Psi_{\zeta}(z)\Psi_{\zeta}(z') \notag \\
&\quad + \left. \Re\big[\tilde{\omega} \, C D(\tilde{\omega})\Psi_{\zeta}(z)\Psi_{\zeta}(z^\prime)\big]\right|_{\tilde{\omega} = \tilde{\omega}_\zeta} \, , \label{eq:delta-res-withBS}
\end{align}
where we used the identity $\Psi_{\zeta}(z)|_{\tilde{\omega} = \overline{\tilde{\omega}_\zeta}} = \overline{\Psi_{\zeta}(z)|_{\tilde{\omega} = \tilde{\omega}_\zeta}}$. The remaining term $D(\tilde{\omega}_\zeta)$ cannot be expressed analytically, but can be defined implicitly (see Appendix~\ref{apx:calculation-delta-expansion}). 

Finally, the bound state mode solutions will also contribute to the two-point function so that its antisymmetric part still obeys \eqref{eq:initial_conditions_E} and, consequently, the CCRs of the quantum field theory are satisfied. Using all the above information in combination with \eqref{eq:2-pt_modes} and \eqref{eq:delta-resolution}, the two-point function for the putative ground state may be written, for each $\zeta\in[\zeta_*,\pi)$,
\begin{widetext}
\begin{align}
G^+_\zeta(x,x^\prime) &= \lim_{\epsilon \to 0^+} \sum_{k\in\mathbb{Z}} e^{ik\left(\tilde{\phi}-\tilde{\phi}^\prime\right)} \int_0^{\infty} \frac{\dd\tilde{\omega}}{(2\pi)^2} \, e^{-i\tilde{\omega} \left(\tilde{t}-\tilde{t}^\prime-i\epsilon\right)} \frac{\left(A\overline{B}-\overline{A}B\right) C}{|{\cos(\zeta) B-\sin(\zeta)A}|^2} \Psi_{\zeta}(z)\Psi_{\zeta}(z') \notag 
\\
&\quad + i \sum_{k\in\mathbb{Z}} e^{ik\left(\tilde{\phi}-\tilde{\phi}^\prime\right)} \left(e^{-i\tilde{\omega}_{\zeta}(\tilde{t}-\tilde{t}^\prime)} + e^{-i\overline{\tilde{\omega}_{\zeta}}(\tilde{t}-\tilde{t}^\prime)}\right) \Re\big[ C D(\tilde{\omega})\Psi_{\zeta}(z)\Psi_{\zeta}(z^\prime)\big]\big|_{\tilde{\omega} = \tilde{\omega}_\zeta} \, . \label{eq:state_Robin_with_BS}
\end{align}
\end{widetext}

Notice that for $\zeta=\zeta_*$ the two bound state frequencies both coincide with the real value $\tilde{\omega} =0$. In this case the integral over positive $\tilde{\omega}$-frequencies has to be interpreted as a Cauchy principal value for $\tilde{\omega}=0$, while the contribution of the bound state mode solutions is calculated using the Sokhotsky-Plemelj formula for distributions.

To conclude this section we comment on the physical significance of the two-point functions obtained in \eqref{eq:state_Robin_no_BS} and \eqref{eq:state_Robin_with_BS}. In the first case, we are dealing with a generalization of \eqref{eq:state_Dirichlet_v2} to Robin boundary conditions. Hence, \eqref{eq:state_Robin_no_BS} is a genuine ground state built only out of positive $\tilde{\omega}$-frequencies and, using once more the results of \cite{Sahlmann:2000fh}, it satisfies the local Hadamard condition. On the contrary, in \eqref{eq:state_Robin_with_BS} there is an additional contribution due to bound state frequencies $\tilde{\omega}_\zeta$, whose existence spoils the property of $G^+_\zeta$ of being a ground state. For this reason, it is not possible to conclude directly whether, in presence of bound sate frequencies, we have constructed a Hadamard, hence physically satisfactory, state. We plan to investigate this issue in future work.

% CONCLUSIONS

\section{Conclusions}
\label{sec:conclusions}

In this paper we have addressed two different, albeit related, questions. The first concerns the structural properties of a real, massive scalar field in BTZ spacetime, with an arbitrary coupling to scalar curvature. More precisely, since the underlying background is not globally hyperbolic, the equation of motion ruling the dynamics cannot be solved only assigning initial data on a partial Cauchy surface (a codimension $1$, acausal, spacelike surface that is intersected by any complete timelike curve at most once), but also a boundary condition at infinity has to be imposed. In this work, we focused our attention on those of Robin type, proving under which constraints on the parameters of the theory they can be imposed, and subsequently constructing explicitly the associated solutions of the equation of motion. 

In the second part of the paper, we have used this result to address whether it is possible to associate to a real, massive scalar field in BTZ spacetime a two-point function, which can be in turn read as the building block of a ground state. We have given a positive and explicit answer to this query for a large class of Robin boundary conditions. Nonetheless, we have highlighted that there exists of a range of boundary conditions that must be excluded, those for which bound state mode solutions occur. When this is not the case, the two-point function possesses some nice physical properties, the most notable one that of being of local Hadamard form. Hence, the states that we have constructed are suitable for defining an algebra of Wick polynomials which are the key ingredient to discuss interactions at a perturbative level. 

Besides offering one of the first examples of a ground state for a quantum field theory in the exterior region of a rotating black hole, this work prompts several future directions of investigation. On the one hand, one could prove the existence of a thermal counterpart of our ground states, hence obtaining in this framework the analogue of the Hartle-Hawking state in Schwarzschild spacetime. On the other hand, one could investigate Hawking radiation in this context and its interplay with the rotation of the black hole, by using the method of Parikh and Wilczek \cite{Parikh:1999mf}, recently extended to the framework of algebraic quantum field theory in \cite{Moretti:2010qd}. A more long term and ambitious goal is the explicit construction of a regularized stress-energy tensor, to be used in the analysis of the semiclassical Einstein's equations, extending the work of \cite{Binosi:1998yu}. We hope to come back to these problems in the near future.

% ACKNOWLEDGEMENTS

\begin{acknowledgments}
	We are grateful to Valter Moretti and Nicola Pinamonti for enlightening discussions. We also thank Nicol\'o Drago, Felix Finster, Klaus Fredenhagen, Alan Garbarz, Carlos Herdeiro, Jorma Louko and Elizabeth Winstanley for comments and discussions. The work of F.~B.\ was supported by a Ph.D. fellowship of the University of Pavia, that of C.~D.\ was supported by the University of Pavia. The work of H.~F.\ was supported by the INFN postdoctoral fellowship ``Geometrical Methods in Quantum Field Theories and Applications'' and by the ``Progetto Giovani GNFM 2017 -- Wave propagation on lorentzian manifolds with boundaries and applications to algebraic QFT'' fostered by the National Group of Mathematical Physics (GNFM-INdAM). The work of I.~K.\ was in part supported by the National Group of Mathematical Physics (GNFM-INdAM).
\end{acknowledgments}

% APPENDICES

\appendix

\section{Delta function as an expansion in eigenfunctions of a quadratic eigenvalue problem}
\label{apx:deltaexpansion}

Our goal in this appendix to give a formula for the expansion of the
delta distribution in terms of eigenfunctions of a differential operator
with quadratic dependence on the spectral parameter like
in~\eqref{eq:Sturm_Liouville_form}, as it is necessary in the
calculations of Section~\ref{Sec:ground_state}. More precisely, we want
to obtain the spectral resolution of the identity for quadratic operator
pencils, specifically concentrating on the case of unbounded operators
coming from Sturm-Liouville ODEs as the one above. While the spectral
theory of polynomial operator pencils has been widely studied
\cite{keldysh,markus}, it is not a topic often covered in
standard references on spectral theory \cite{reedsimon,kato}.

Consider a family of operators defined on a Hilbert space $\H$, referred
to as a \emph{quadratic operator pencil},
\begin{equation}
S_\tomega = P + \tomega \cR_1 + \tomega^2 \cR_2 \, ,
\end{equation}
with (S1) $\cR_1$, $\cR_2$ and $\cR_2^{-1}$ all bounded and self-adjoint,
and $P$ unbounded, closed and hermitian on a dense domain $D(S_\tomega) \subset
\H$, as is the case with our main example $S_\tomega = \cJ^{-1}
L_\tomega$ on $\H = L^2((0,1); \cJ(z)\,\dd{z})$, where $L_\tomega$ is
defined in~\eqref{eq:Sturm_Liouville_form} and $\cJ(z)$
in~\eqref{eq:measure}.

\begin{figure}
	\begin{center}
		\includegraphics[scale=1.25]{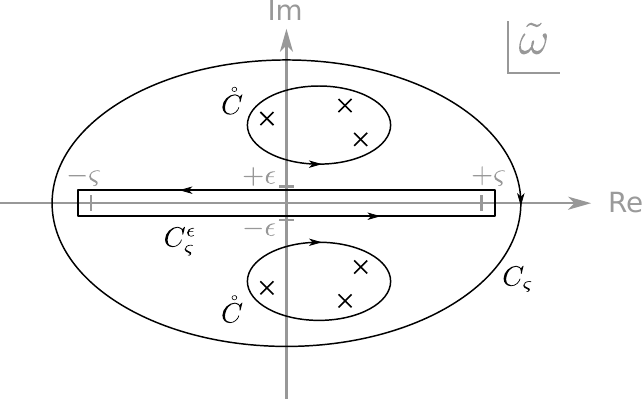}
	\end{center}
	\caption{Contour for the integral representation of the identity
		operator in~\eqref{eq:res-ident}.}
	\label{fig:contour}
\end{figure}

Define the \emph{resolvent} of $S_\tomega$ as $T_\tomega = S_\tomega^{-1}$,
when it exists. The \emph{resolvent set} $\rho(S_\tomega) \subset \bC$ consists
of all values of $\tomega \in \bC$ such that $T_\tomega$ exists and is a
bounded operator. As usual, we define the \emph{spectrum} $\sigma(S_\tomega) =
\bC \setminus \rho(S_\tomega)$. We will show that, when (S2) $\sigma(S_\tomega)$
consists only of a subset of $\bR$ together with a finite number of
isolated points in $\bC \setminus \bR$ symmetric with respect to complex
conjugation, the identity operator can be represented by the
integral
\begin{align} \label{eq:res-ident}
I &= \lim_{\varsigma\to\oo} \int_{-\varsigma}^\varsigma
\frac{\dd\tomega}{2\pi i} \,
\lim_{\epsilon\to 0^+} \tomega (T_{\tomega-i\epsilon} - T_{\tomega+i\epsilon}) \cR_2 \notag \\
&\quad + \oint_{\mathring{C}} \frac{\dd\tomega}{2\pi i} \tomega T_\tomega \, ,
\end{align}
where the contour $\mathring{C}$ illustrated in Figure~\ref{fig:contour}
positively and simply encircles the non-real part of the spectrum, the
inner $\epsilon \to 0^+$ limit is taken in the sense of distributions in
$\tomega$ (boundary values of holomorphic functions define a special
kind of distribution~\cite[Ch.IX]{hoermander}) and the outer $\varsigma
\to \oo$ limit is taken in the sense of the strong operator topology.

The key idea is to \emph{linearize} the quadratic operator pencil to a
linear operator pencil $\bS_\tomega$, while doubling the size of the
Hilbert space, in a way that keeps the spectral problems of $S_\tomega$
and $\bS_\tomega$ equivalent. Since the spectral theory of linear operator
pencils (essentially, generalized eigenvalue problems) is well known, we
can leverage this equivalence to obtain the desired formulas for
$S_\tomega$. More precisely, consider the following linear operator
pencil defined on $\H^2 = \H \oplus \H$,
\begin{equation}\label{eq:linearpencil}
\bS_\tomega
= \P + \tomega \R
= \begin{bmatrix}
P & 0 \\
0 & -\cR_2
\end{bmatrix}
+ \tomega \begin{bmatrix}
\cR_1 & \cR_2 \\
\cR_2 & 0
\end{bmatrix} .
\end{equation}
The linear pencil $\bS_\tomega$ is related to the quadratic one $S_\tomega$
by the basic identities
\begin{align}
\bS_\tomega
\begin{bmatrix} I \\ \tomega \end{bmatrix} \Psi
&= \begin{bmatrix} I \\ 0 \end{bmatrix} S_\tomega \Psi \, ,
\\
S_\tomega \begin{bmatrix} I & 0 \end{bmatrix}
\begin{bmatrix} \Psi \\ \Phi \end{bmatrix}
&= \begin{bmatrix} I & \tomega \end{bmatrix}
\bS_\tomega
\begin{bmatrix} \Psi \\ \Phi \end{bmatrix} .
\end{align}
It is easy to see that, when $\cR_1$ and $\cR_2$ are bounded and
self-adjoint, so is $\R$, and when in addition $P$ is closed and
hermitian on $D(S_\tomega)$, so is $\P$ on $D(\bS_\tomega) = D(S_\tomega) \oplus \H$. While,
there are many possible linearizations of a quadratic operator pencil,
we have chosen this one to preserve these self-adjointness properties. 

Define the resolvent $\T_\tomega = \bS_\tomega^{-1}$, when it
exists. The spectrum and resolvent set $\sigma(\bS_\tomega), \, \rho(\bS_\tomega) \subset
\bC$ are defined in the usual way, essentially exactly as above. Direct
calculation shows that, when both exist, the resolvents of $S_\tomega$
and $\bS_\tomega$ are related to each other by
\begin{align}
\notag
\T_\tomega
&= \begin{bmatrix} I \\ \tomega \end{bmatrix}
T_\tomega
\begin{bmatrix} I & \tomega \end{bmatrix}
+ \begin{bmatrix} 0 & 0 \\ 0 & -\cR_2^{-1} \end{bmatrix} \\
&= \begin{bmatrix}
T_\tomega & \tomega T_\tomega \\
\tomega T_\tomega & \tomega^2 T_\tomega - \cR_2^{-1}
\end{bmatrix}
,
\label{eq:quadR2linR}
\end{align}
\begin{align}
T_\tomega
&= \begin{bmatrix} I & 0 \end{bmatrix}
\T_\tomega
\begin{bmatrix} I \\ 0 \end{bmatrix}
= \frac{1}{\tomega} \begin{bmatrix} I & 0 \end{bmatrix}
\T_\tomega
\begin{bmatrix} 0 \\ I \end{bmatrix}
.
\end{align}
From the above formulas it is clear that when $T_\tomega$ exists and is
bounded, so is $\T_\tomega$, and vice-versa. Thus $\rho(S_\tomega) =
\rho(\bS_\tomega)$ and, necessarily, $\sigma(S_\tomega) = \sigma(\bS_\tomega)$, which makes
precise the sense in which the spectral problems of the two operator
pencils equivalent. Once we know what $\rho(S_\tomega)$ is, using the
boundedness of $\R$, a variant of Theorem~VI.5 of~\cite{reedsimon} shows that
$\T_\tomega$ is analytic on $\rho(S_\tomega)=\rho(\bS_\tomega)$, which implies by the
explicit relationship between them that $T_\tomega$ is also analytic on
$\rho(S_\tomega)$.

Let $\nu \in \rho(\bS_\tomega) = \rho(S_\tomega)$ and let $C_\nu \subset \rho(\bS_\tomega) =
\rho(S_\tomega)$ be a contour that simply encircles $\nu$, though in the
negative direction, meaning that, upon deformation, $C_\nu$ has a chance
of simply and positively encircling $\sigma(S_\tomega)$. Though, since our
$\sigma(S_\tomega)$ is unbounded, the deformation of the contour will have to go
through a limiting procedure. There is no need for $C_\nu$ to be
connected. In fact, it is advantageous to have a connected component of
$C_\nu$ contained in each connected component of $\rho(S_\tomega)$. Provided
that the resolvent $\G_\tomega$ is analytic on $\rho(S_\tomega)$, the Cauchy
residue formula gives
\begin{equation}
\T_\nu \R = -\oint_{C_\nu} \frac{\dd\tomega}{2\pi i} \frac{1}{\tomega -\nu} \T_\tomega \R \, .
\end{equation}
Multiplying both sides by $\R^{-1}\bS_\nu$, we get
\begin{equation} \label{eq:cauchy-Cnu}
\mathbf{I} = \oint_{C_\nu} \frac{\dd\tomega}{2\pi i}
\left(
\T_\tomega \R - \frac{\mathbf{I}}{\tomega-\nu}
\right) ,
\end{equation}
where the contour $C_\nu$ can be deformed at will, as long as it remains
within $\rho(S_\tomega) \setminus \{\nu\}$.

We can deform the contour $C_\nu$ to the desired limiting form
in~\eqref{eq:res-ident} if we can take advantage of an abstract spectral
representation for the operator $\R^{-1}\P$, that is (S3) there exists a
projection operator valued measure $\E(\nu)$ on $\sigma(\bS_\tomega)$,
satisfying the usual commutation and monotonicity conditions, giving the
spectral representation $\R^{-1}\P = \int_{\sigma(\bS_\tomega)} \nu \, \dd \E(\nu)$.
As a consequence, we also get the spectral representation $\T_\tomega \R =
\int_{\sigma(\bS_\tomega)} \frac{1}{\nu+\tomega} \dd \E(\nu)$. If we let
$\E_\varsigma = \E(\{\nu \in \bC \mid |\nu|<\varsigma\})$, then
$\E_\varsigma \to \mathbf{I}$ strongly as $\varsigma \to \oo$ and
$\bigcup_{\varsigma>0} \operatorname{ran} \E_\varsigma $ is dense in $\H^2$.

Another consequence of the abstract spectral representation is that
$\T_\tomega \R \E_\varsigma$ is now analytic for $|\tomega| > \varsigma$
and has the strong asymptotic expansion $\T_\tomega \R \E_\varsigma =
\frac{1}{\tomega} \E_\varsigma + \mathcal{O}(\frac{1}{\tomega^{2}})$. Multiplying
both sides of~\eqref{eq:cauchy-Cnu} by $\E_\varsigma$ we get
\begin{equation*}
\E_\varsigma = \oint_{C_\nu} \frac{\dd\tomega}{2\pi i}
\left( \E_\varsigma \T_\tomega \R - \frac{\E_\varsigma}{\tomega-\nu} \right)
- \oint_{C_\nu} \frac{\dd\tomega}{2\pi i} \frac{\mathbf{I}-\E_\varsigma}{\tomega-\nu} \, .
\end{equation*}
The second integral can be evaluated immediately and combined with the
left-hand side. In the first integral, we can deform the contour $C_\nu$
to the contour $C_\varsigma \cup C^\epsilon_\varsigma \cup
\mathring{C}$, as illustrated in Figure~\ref{fig:contour}. Because
the asymptotics mentioned above, the integral over the large circle
$C_\varsigma$ contributes at the order $\mathcal{O}(\frac{1}{\varsigma})$. On the
other hand, the term $\frac{\E_\varsigma}{\tomega-\nu}$ is analytic over
the contours $\mathring{C}$, $C^\epsilon_\varsigma$ and their interiors,
so its contribution vanishes, which leaves us with
\begin{equation} \label{eq:cauchy-Ceps}
\mathbf{I}
= \oint_{C^\epsilon_\varsigma} \frac{\dd\tomega}{2\pi i} \,
\T_\tomega \R \E_\varsigma
+ \oint_{\mathring{C}} \frac{\dd\tomega}{2\pi i} \,
\T_\tomega \R \E_\varsigma
+ \mathcal{O}(\varsigma^{-1}) .
\end{equation}

Next, before taking the limits $\epsilon \to 0^+$ and $\varsigma \to
\oo$, we multiply both sides of~\eqref{eq:cauchy-Ceps} by an arbitrary
$\mathbf{v}_{\varsigma'} \in \H^2$ such that $\mathbf{v}_{\varsigma'} =
\E_{\varsigma} \mathbf{v}_{\varsigma'}$ for any $\varsigma >
\varsigma'$, so that
\begin{align*}
\mathbf{v}_{\varsigma'}
&= \oint_{C^\epsilon_\varsigma} \frac{\dd\tomega}{2\pi i} \,
\T_\tomega \R \E_\varsigma \mathbf{v}_{\varsigma'}
+ \oint_{\mathring{C}} \frac{\dd\tomega}{2\pi i} \,
\T_\tomega \R \E_\varsigma \mathbf{v}_{\varsigma'}
+ \mathcal{O}(\varsigma^{-1}) \\
&= \int_{-\varsigma}^\varsigma \frac{\dd\tomega}{2\pi i} \,
\lim_{\epsilon\to 0^+} (\T_{\tomega-i\epsilon} -
\T_{\tomega+i\epsilon}) \R \E_\varsigma \mathbf{v}_{\varsigma'} \\
& \quad {}
+ \oint_{\mathring{C}} \frac{\dd\tomega}{2\pi i} \,
\T_\tomega \R \E_\varsigma \mathbf{v}_{\varsigma'}
+ \mathcal{O}(\varsigma^{-1}) \\
&= \lim_{\varsigma\to \oo} \int_{-\varsigma}^\varsigma
\frac{\dd\tomega}{2\pi i} \lim_{\epsilon\to 0^+} 
(\T_{\tomega-i\epsilon} - \T_{\tomega+i\epsilon}) \R \mathbf{v}_{\varsigma'} \\
& \quad {}
+ \oint_{\mathring{C}} \frac{\dd\tomega}{2\pi i} \,
\T_\tomega \R \mathbf{v}_{\varsigma'} \, .
\end{align*}
Note that the $\epsilon \to 0^+$ limit is taken in the distributional
sense with respect to $\tomega$. Finally, using a variant of the
Banach-Steinhaus theorem (Theorem~2.11.4 of ~\cite{HillePhillips}), we obtain the
following strong limit
\begin{equation*} %\label{eq:lin-res-ident-reg}
\R^{-1}
= \lim_{\varsigma \to \oo} \int_{-\varsigma}^\varsigma
\frac{\dd \tomega}{2\pi i} \,
\lim_{\epsilon\to 0^+} (\T_{\tomega-i\epsilon} - \T_{\tomega+i\epsilon})
+ \oint_{\mathring{C}} \frac{\dd\tomega}{2\pi i} \, \T_\tomega \, ,
\end{equation*}
where to apply the theorem we need to recall that finite linear
combinations of vectors like $\mathbf{v}_{\varsigma'}$ are dense in
$\H^2$ and note that due to~\eqref{eq:cauchy-Ceps} the norms of the
integrals
\[
\int_{-\varsigma}^\varsigma \frac{\dd \tomega}{2\pi i}\,
\lim_{\epsilon\to 0^+} (\T_{\tomega-i\epsilon} - \T_{\tomega+i\epsilon})
+ \oint_{\mathring{C}} \frac{\dd\tomega}{2\pi i} \, \T_\tomega
\]
are uniformly bounded for large $\sigma$.

Using the second equality in~\eqref{eq:quadR2linR} and the formula
\begin{equation}
\R^{-1}
= \begin{bmatrix}
\cR_1 & \cR_2 \\
\cR_2 & 0
\end{bmatrix}^{-1}
= \begin{bmatrix}
0 & \cR_2^{-1} \\
\cR_2^{-1} & -\cR_2^{-1} \cR_1 \cR_2^{-1}
\end{bmatrix}
\end{equation}
finally gives us the desired identity~\eqref{eq:res-ident}.

The argument we have just presented, for the linear operator pencil,
mimicks that of~\cite[Ch.9]{weidmann}. There, the existence of the
spectral measure $\E(\nu)$ followed from the standard spectral theorem
for self-adjoint operators on a Hilbert space, with the operator
$\R^{-1} \P$ being self-adjoint with respect to the weighted inner
product $[\mathbf{v},\mathbf{u}] = (\mathbf{v}, \R \mathbf{u})$, which
was assumed to be positive definite. In our case,
$[\mathbf{v},\mathbf{u}]$ is clearly indefinite and thus defines a Krein
space $\cK = (\H^2, [-,-])$ rather than a Hilbert space. Fortunately, in
the Krein space setting we can still appeal to a spectral theorem,
provided that the operator $\R^{-1} \P$ is definitizable. This will
indeed be the case for the specific operators defined in
Appendix~\ref{apx:calculation-delta-expansion}. Though, since verifying the necessary hypothesis
is rather technical, we relegate them to
Appendix~\ref{apx:S3}. A more hands-on alternative to
hypothesis (S3) would be a direct estimate of the form $\T_\tomega =
\frac{1}{\tomega} \R^{-1} + \mathcal{O}(\frac{1}{\tomega^2})$ that is uniform over
a neigborhood of $\tomega = \oo$ minus a sector of positive angle
containing the real axis. Such an estimate could be obtained by a WKB
analysis of the differential operators discussed in Appendix~\ref{apx:calculation-delta-expansion},
which may be considered in future work.

\vspace*{5ex}

\section{Explicit calculation of the delta integral representation}
\label{apx:calculation-delta-expansion}

In this appendix, we show in detail the procedure to compute the delta integral representation \eqref{eq:delta-resolution} and, hence, the mode expansion of the two-point function \eqref{eq:2-pt_modes} for the case in which the mass parameter is such that $-1 < \mu^2 < 0$ and Robin boundary conditions parametrized by $\zeta \in [0,\pi)$ are imposed at $z=1$. The results for $\mu^2 \geqslant 0$ may be simply obtained by setting $\zeta = 0$.

Now, let us apply the general discussion from
Appendix~\ref{apx:deltaexpansion} to the differential operator $L_\tomega$
introduced in~\eqref{eq:Sturm_Liouville_form}, which we write for convenience as
\begin{multline}
	L_\tomega\Psi(z)
	= \frac{\dd}{\dd z} \left( z \frac{\dd \Psi(z)}{\dd z} \right)
	- \bigg[ \frac{\ell^2 k^2(1-z)-r_+^2\mu^2}{4 r_+^2(1-z)} 
	\\
	- \frac{\tilde{\omega} \ell^3 k r_-}{2r_+ (r_+^2-r_-^2) (1-z)}
	- \frac{\tilde{\omega}^2\ell^4 \cJ(z)}{4(r_+^2-r_-^2)}
	\bigg] \Psi(z)	, \label{eq:B1}
\end{multline}
with $\cJ(z)$ the same as in~\eqref{eq:measure}. We let the Hilbert
space be $\H = L^2((0,1); \cJ(z)\dd{z})$ and we let the quadratic operator pencil be
\begin{equation} \label{eq:SLop}
S_{\tilde{\omega}}\Psi(z) = \frac{1}{\cJ(z)} L_\tomega\Psi(z) .
\end{equation}
This operator satisfies the hypotheses (S1), (S2) and (S3) from
Appendix~\ref{apx:deltaexpansion}. The verification of the hypotheses is of a much more technical nature and is relegated to Appendix~\ref{apx:S1}, 
Appendix~\ref{apx:S2} and Appendix~\ref{apx:S3}, respectively.

We want to construct a Green's distribution $\cG_{\tomega,\zeta}$ associated to $L_\tomega$ 
consisting of the product of square integrable solutions of $L_\tomega \Psi = 0$ at both $z=0$ and $z=1$.
For that we introduce
\begin{equation}
u_{\tilde{\omega}}(z) = \begin{cases}
\Psi_3(z) \, , & \Im[\tilde{\omega}] > 0 \, , \\
\Psi_4(z) \, , & \Im[\tilde{\omega}] < 0 \, ,
\end{cases} \label{eq:functionu}
\end{equation}
with $\Psi_3$ and $\Psi_4$ defined in \eqref{Psi0}, which is uniquely chosen by the property of being
$L^2$ at $z=0$, as seen in Section~\ref{sec:endpoints}. 
We also introduce
\begin{equation} \label{eq:functionv}
\Psi_{\tomega,\zeta}(z) = \cos(\zeta)\Psi_1(z)+\sin(\zeta)\Psi_2(z) \, ,
\end{equation}
with $\Psi_1$ and $\Psi_2$ defined either
by~\eqref{eq:z1gen} or~\eqref{eq:z1spec}, which is uniquely chosen by the property of being
$L^2$ at $z=1$  when $-1<\mu^2<0$ and satisfying Robin boundary conditions parametrized by $\zeta\in [0,\pi)$.
Note that, given the identity $\overline{L_{\tomega}} =
L_{\overline{\tomega}}$, one has
$u_{\overline{\tomega}} = \overline{u_{\tomega}}$ and
$\Psi_{\overline{\tomega},\zeta} = \overline{\Psi_{\tomega,\zeta}}$.

The Green's distribution $\cG_{\tomega,\zeta}$ may then be written as
\begin{equation}
\cG_{\tomega,\zeta}(z,z')
= 
\begin{cases}
\mathcal{N}^{-1}_{\tomega,\zeta}\ u_\tomega(z) \Psi_{\tomega,\zeta}(z') \ , & z\leqslant z' \ , \\
\mathcal{N}^{-1}_{\tomega,\zeta}\ u_\tomega(z') \Psi_{\tomega,\zeta}(z) \ , & z\geqslant z' \ ,
\end{cases} 
\label{eq:radial_Green_BC}
\end{equation}
with
\begin{widetext}
	\begin{align}
	\mathcal{N}_{\tilde{\omega},\zeta} = - z \mathcal{W}_z \left[u_{\tilde{\omega}},\Psi_{\zeta}\right]
	= \begin{cases} 
	\cos(\zeta) \, \dfrac{\Gamma(c)\Gamma(a+b-c+1)}{\Gamma(a)\Gamma(b)} + \sin(\zeta) \, \dfrac{\Gamma(c)\Gamma(c-a-b+1)}{\Gamma(c-a)\Gamma(c-b)} \, , & \Im[\tilde{\omega}] > 0 \, , \\
	\cos(\zeta) \, \dfrac{\Gamma(2-c)\Gamma(a+b-c+1)}{\Gamma(a-c+1)\Gamma(b-c+1)} + \sin(\zeta) \, \dfrac{\Gamma(2-c)\Gamma(c-a-b+1)}{\Gamma(1-a)\Gamma(1-b)} \, , & \Im[\tilde{\omega}] < 0 \, ,
	\end{cases} \label{eq:Wronski_BC}
	\end{align}
\end{widetext}
where the parameters $a,b,c$ are as in \eqref{eq:parameters}. The normalization constant $\mathcal{N}_{\tilde{\omega},\zeta}$ was evaluated using the intermediate result
$$\mathcal{W}_z[\Psi_1,\Psi_2] = \frac{a+b-c}{z} = \frac{\sqrt{1+\mu^2}}{z} \, , $$
and the following connection formulas of hypergeometric functions (see Eqs.~(15.10.17-18) of \cite{NIST}):
\begin{subequations}
	\begin{align} 
	\Psi_3(z) &=
	\frac{\Gamma(c)\Gamma(c-a-b)}{\Gamma(c-a)\Gamma(c-b)}\Psi_1(z) \notag \\ 
	&\quad +\frac{\Gamma(c)\Gamma(a+b-c)}{\Gamma(a)\Gamma(b)}\Psi_2(z) \, , \label{eq:connection3} \\
	\Psi_4(z) &= \frac{\Gamma(2-c)\Gamma(c-a-b)}{\Gamma(1-a)\Gamma(1-b)}\Psi_1(z) \notag \\ 
	&\quad +\frac{\Gamma(2-c)\Gamma(a+b-c)}{\Gamma(a-c+1)\Gamma(b-c+1)}\Psi_2(z) \, . \label{eq:connection4}
	\end{align}
\end{subequations}	

By inspection of~\eqref{eq:radial_Green_BC} and~\eqref{eq:Wronski_BC}, 
one has that  $\overline{\mathcal{N}_{\tomega,\zeta}} =
\mathcal{N}_{\overline{\tomega},\zeta}$ and $\overline{\cG_{\tomega,\zeta}}(z,z') 
= \cG_{\overline{\tomega},\zeta}(z',z)$. 
Moreover, as noted in Appendix~\ref{apx:S2}, $\mathcal{N}_\tomega$ is 
analytic on $\Im[\tomega] \ne 0$ and has at most two isolated 
zeros, the \emph{bound state frequencies}, that are reflection symmetric about the real axis,
forming a set $\BS_\zeta \subset \bC$ such that
$\BS_\zeta = \BS_\zeta^+ \cup \overline{\BS_\zeta^+}$ with
$\Im[\BS_\zeta^+] > 0$.

We can now apply formula
\eqref{eq:res-ident} to write the following integral
representation of the delta distribution:
\begin{align}
\frac{4(r_+^2 - r_-^2)}{\ell^4 \cJ(z)} \delta(z-z')
& = -\int_{\bR} \frac{\dd\tilde{\omega}}{2\pi i} \,
\tilde{\omega} \,
\Delta \cG_{\tomega,\zeta}(z,z') \notag
\\
&\quad + \oint_{\mathring{C}} \frac{\dd\tomega}{2\pi i} \,
\tomega \mathcal{G}_{\tomega,\zeta}(z,z') \, ,
\end{align}
where the contour $\mathring{C}$ illustrated in Figure~\ref{fig:contour}
positively and simply encircles the bound state frequencies in $\BS_\zeta$,
and
\begin{equation}
\Delta \cG_{\tomega,\zeta}(z,z')
\doteq \lim_{\epsilon \to 0^+}
[\cG_{\tomega+i\epsilon,\zeta}(z,z') - \cG_{\tomega-i\epsilon,\zeta}(z,z')]
\end{equation}
should be interpreted as a distribution in $\tomega$.
An application of Cauchy's residue theorem
gives the integral representation
\begin{align}
\frac{\delta(z-z')}{\cJ(z)}
\notag
&= -\frac{\ell^4}{4(r_+^2-r_-^2)} \Bigg[ \int_\bR \frac{\dd\tomega}{2\pi i} \tomega \,
\Delta \cG_{\tomega,\zeta}(z,z') \\
& \quad
+ \sum_{\tomega'\in \BS_\zeta}
\Res_{\tomega=\tomega'}[ \tomega \, \cG_{\tomega,\zeta}(z,z')] \Bigg] \ ,
\label{eq:id-res}
\end{align}

Both integrands in~\eqref{eq:id-res}
can be computed rather explicitly, except for analytic expressions for the
bound state frequencies (see Appendix~\ref{apx:S2}). 
Introducing
\begin{equation*}
A=\frac{\Gamma(c-1)\Gamma(c-a-b)}{\Gamma(c-a)\Gamma(c-b)} , \quad B=\frac{\Gamma(c-1)\Gamma(a+b-c)}{\Gamma(a)\Gamma(b)} ,
\end{equation*}
and using the connection formulas \eqref{eq:connection3} and \eqref{eq:connection4}, one may write
\begin{equation*}
u_{\tilde{\omega}}(z) = \begin{cases}
(c-1) \left[A \Psi_1(z) + B \Psi_2(z) \right] \, , & \Im[\tilde{\omega}] > 0 \, , \\
(1-c) \left[\overline{A} \Psi_1(z) + \overline{B} \Psi_2(z) \right] \, , & \Im[\tilde{\omega}] < 0 \, ,
\end{cases}
\end{equation*}
and
\begin{align*}
\mathcal{N}_{\tilde{\omega},\zeta}
= \begin{cases} 
(1-c) \sqrt{1+\mu^2} \big[{\cos(\zeta)} B - \sin(\zeta) A \big], & \Im[\tilde{\omega}] > 0 , \\
(c-1) \sqrt{1+\mu^2} \big[{\cos(\zeta)} \overline{B} - \sin(\zeta) \overline{A} \big], & \Im[\tilde{\omega}] < 0 .
\end{cases}
\end{align*}
Hence, for $z < z'$, 
\begin{align}
\Delta \mathcal{G}_{\tilde{\omega}}(z,z')
&= - \frac{1}{\sqrt{1+\mu^2}} \left[ \frac{A \Psi_1(z) + B \Psi_2(z)}{\cos(\zeta) B - \sin(\zeta) A} \right. \notag \\
&\quad \left. - \frac{\overline{A} \Psi_1(z) + \overline{B} \Psi_2(z)}{\cos(\zeta) \overline{B} - \sin(\zeta) \overline{A}} \right] \Psi_{\zeta}(z') \notag
\\
&= \frac{\overline{A}B-A\overline{B}}{\left|{\cos(\zeta) B-\sin(\zeta) A}\right|^2} \frac{\Psi_{\zeta}(z)\Psi_{\zeta}(z')}{\sqrt{1+\mu^2}} \, ,
\end{align}
and the result is also valid for $z > z'$.

Now, let us consider the residues at a bound state frequency $\tomega_\zeta
\in \BS^+_\zeta$. When it exists, it is an isolated root of
$\mathcal{N}_{\tomega,\zeta} = 0$ and
\begin{equation}
\Res_{\tomega=\tomega_\zeta} [\tomega \cG_{\tomega,\zeta}(z,z')]
= \frac{\tomega_\zeta}{2} D(\tomega_\zeta)
\Psi_{\tomega_\zeta,\zeta}(z) \Psi_{\tomega_\zeta,\zeta}(z') ,
\end{equation}
where $D(\tomega_\zeta) = D_2(\tomega_\zeta)/D_1(\tomega_\zeta)$. From the Laurent series
of $\mathcal{N}_{\tomega,\zeta}$ we get
\begin{align*}
D_1(\tilde{\omega_\zeta}) 
&\doteq \frac{\ell^2 \sqrt{1+\mu^2}}{i(r_+^2-r_-^2)}  \, \big\{{\sin(\zeta)}\, A \big[(r_++r_-)\psi(c-a) \\
&\qquad +(r_+-r_-)\psi(c-b)-2r_+\psi(c) \big] (1-c) \\
&\quad - \cos(\zeta)\,B \big[(r_++r_-)\psi(b)+(r_+-r_-)\psi(a) \\
&\qquad -2r_+\psi(c)\big] (1-c)\big\}|_{\tomega=\tomega_\zeta} \, ,
\end{align*}
where $\psi$ is the digamma function. Since
$\mathcal{N}_{\tomega_\zeta,\zeta} = 0$, the solutions $u_{\tomega_\zeta}$ and
$\Psi_{\tomega_\zeta,\zeta}$ are no longer linearly independent and their ratio (recall that $\Im[\tomega_\zeta] > 0$) is
\begin{align*}
D_2(\tilde{\omega_\zeta})
&\doteq \frac{u_{\tomega_\zeta}(z)}{\Psi_{{\tomega_\zeta},\zeta}(z)}\\ 
&= \begin{cases}
\sec(\zeta) (c-1) A|_{\tomega=\tomega_{\zeta}}  \, , & \cos(\zeta) \ne 0 \, , \\
\csc(\zeta) (c-1) B|_{\tomega=\tomega_{\zeta}}  \, , & \sin(\zeta) \ne 0 \, .
\end{cases}
\end{align*}
%
% We call the corresponding $\Psi_{\tomega_{\zeta},\zeta}(z)$ a \emph{bound state mode solution}.

Finally, the spectral resolution of the delta distribution takes the form
\begin{multline}
	\frac{\delta(z-z^\prime)}{\cJ(z)} 
	= \frac{\ell^4}{4(r_+^2-r_-^2)} 
	\\
	\times \Bigg[ \int_{\bR} \frac{\dd\tilde{\omega}}{2\pi i} \, \tilde{\omega} \, \frac{A\overline{B}-\overline{A}B}{|{\cos(\zeta) B-\sin(\zeta) A}|^2} \frac{\Psi_\zeta(z)\Psi_\zeta(z^\prime)}{\sqrt{1+\mu^2}}
	\\
	+ \sum_{\tomega_{\zeta} \in \BS^+_{\zeta}} \Re\big[ \tilde{\omega}_{\zeta} D(\tilde{\omega}_{\zeta})\Psi_{\tomega_{\zeta},\zeta}(z)\Psi_{\tomega_{\zeta},\zeta}(z^\prime)\big] \Bigg] \, .
\end{multline}
We have taken advantage of the fact that bound state
frequencies come in complex conjugate pairs, $\BS_{\zeta} = \BS^+_{\zeta} \cup
\overline{\BS^+_{\zeta}}$, and of the identities $\overline{D(\tomega_{\zeta})} =
D(\overline{\tomega_{\zeta}})$, $\overline{\Psi_{\tomega_{\zeta},\zeta}(z)} =
\Psi_{\overline{\tomega_{\zeta}},\zeta}(z)$.

\section{Check of hypothesis (S1)}
\label{apx:S1}

In this appendix we show that hypothesis (S1) of Appendix \ref{apx:deltaexpansion} is verified for the quadratic operator pencil $S_{\tilde{\omega}}$.

First, we discuss the relation of the domain of $S_{\tilde{\omega}}$, 
$D(S_{\tilde{\omega}})\subset\mathcal{H} = L^2((0,1); \cJ(z)\dd{z})$, to the choice of
boundary conditions for $L_\tomega$ in \eqref{eq:SLop}. By standard
arguments~\cite[Ch.3]{weidmann}, each choice of boundary conditions will
give a closed operator realization of $S_\tomega$ on a dense domain
$D(S_\tomega)$. Then, if there exists at least one $\tomega \in \bC$ such that
$\tomega,\overline{\tomega} \in \rho(S_\tomega)$ and the corresponding bounded
resolvents satisfy $T_\tomega^* = T_{\overline{\tomega}}$, the closed
operator $S_\tomega$ will be self-adjoint, in the sense that $S^*_\tomega =
S_{\overline{\tomega}}$ and $D(S_\tomega^*) = D(S_\tomega)$.
Hence, we need to check that (a) the Green's distribution associated to $L_\tomega$, $\cG_\tomega$, exists for at
least one $\tomega \in \bC$, that (b) $T_\tomega = \cG_\tomega \cJ$ is
bounded for at least one $\tomega \in \bC$ and that (c) we can satisfy
$\overline{\cG_\tomega}(z,z') = \cG_{\overline{\tomega}}(z',z)$ and
hence $T^*_{\tomega} = T_{\overline{\tomega}}$. The properties (a)
and (c) are explicitly checked in Appendix \ref{apx:calculation-delta-expansion} for each choice of Robin boundary
conditions parametrized by $\zeta$. 

In order to check property (b), 
we need to prove the boundedness of the resolvent $T_\tomega=\mathcal{G}_\tomega\mathcal{J}$. Using the same notation of Appendix~\ref{apx:calculation-delta-expansion}, for a given $\tomega$ with $\Im[\tomega] \ne 0$, provided that $u_\tomega$ and
$\Psi_{\tomega,\zeta}$ introduced in \eqref{eq:functionu} and \eqref{eq:functionv} are linearly independent, that is,
$\mathcal{N}_{\tomega,\zeta}$ in \eqref{eq:Wronski_BC} does not vanish, we can get boundedness starting from the more precise asymptotic estimates:

\begin{subequations} \label{eq:estimates}
	\begin{align}
	|u_\tomega(z)| &\lesssim
	z^{\lambda} (1-z)^{1-\beta-\epsilon} \ , \\
	|\Psi_{\tomega,\zeta}(z)| &\lesssim
	\begin{cases}
	z^{-\lambda} (1-z)^{\beta} \ , & \zeta=0 \ , \\
	z^{-\lambda} (1-z)^{1-\beta-\epsilon} \ , & \zeta\ne0 \ .
	\end{cases}
	\end{align}
\end{subequations}
Here, $\lambda \doteq \ell^2 r_+ \left|\Im\tomega\right| / 2(r_+^2-r_-^2)$ and the symbol $\lesssim$ denotes an
inequality up to a multiplicative constant, uniform over $z\in (0,1)$
where applicable. The constant $\epsilon > 0$ helps to cover the cases with
logarithmic singularities and it could be chosen to depend on other
parameters. Using the same notation, we also have
\begin{equation}
|\cJ(z)| \lesssim z^{-1} (1-z)^{-1} \, .
\end{equation}

The strategy to show boundedness of $T_\tomega = \cG_\tomega \cJ$ is to
apply the so-called \emph{weighted Schur test} (Theorem~5.2 of ~\cite{halmos}).
The inequalities, where, after a factorization, we apply the
Cauchy-Schwarz inequality,
\begin{widetext}
	\begin{align*}
	\|T_\tomega\Psi\|^2
	&= \int_0^1\dd{z} \cJ(z) \left|\int_0^1 \dd{z'} \cG_\tomega(z,z') \cJ(z') \Psi(z') \right|^2 \\
%	&\leqslant \int_0^1\dd{z} \cJ(z) \left(\int_0^1
%	\sqrt{\left|\cG_\tomega(z,z')\right| \cJ(z')\cJ_1(z')}
%	\sqrt{\left|\cG_\tomega(z,z')\right| \frac{\cJ(z')}{\cJ_1(z')}
%		\left|\Psi(z')\right|^2} \dd{z'}\right)^2 \\
	&\leqslant \int_0^1\dd{z} \cJ(z)
	\left(\int_0^1\dd{z'} \left|\cG_\tomega(z,z')\right| \cJ(z') \cJ_1(z') \right)
	\left(\int_0^1\dd{z'}
	\left|\cG_\tomega(z,z')\right| \frac{\cJ(z')}{\cJ_1(z')}
	\left|\Psi(z)\right|^2 \right) 
	\\
	&\leqslant \int_0^1\dd{z'}
	\left(\int_0^1\dd{z} \cJ(z) \cJ_2(z) \left|\cG_\tomega(z,z')\right|\right)
	\frac{\cJ(z')}{\cJ_1(z')} \left|\Psi(z')\right|^2
	\\ 	
	&\leqslant \int_0^1\dd{z'} \frac{\cJ_3(z')}{\cJ_1(z')}
	\cJ(z') \left|\Psi(z')\right|^2 \, ,
	\end{align*}
\end{widetext}
show that $\|T_\tomega \Psi\|^2 \lesssim \|\Psi\|^2$, provided we can
find functions $\cJ_1(z)$, $\cJ_2(z)$, $\cJ_3(z)$ satisfying the
estimates
\begin{gather*}
\int_0^1 \dd{z'} \left|\cG_\tomega(z,z')\right| \cJ(z') \cJ_1(z')
\lesssim \cJ_2(z) \, , \\
\int_0^1 \dd{z} \cJ(z) \cJ_2(z) \left|\cG_\tomega(z,z')\right|
\lesssim \cJ_3(z') \, , \\
\frac{\cJ_3(z')}{\cJ_1(z')}
\lesssim 1 \, .
\end{gather*}
The only free choice is actually in $\cJ_1$, since $\cJ_2$ and $\cJ_3$
(or rather their lower bounds) are then determined by the properties of
$\cG_\tomega(z,z')$. Given the estimates~\eqref{eq:estimates} and
formula~\eqref{eq:radial_Green_BC}, it is straightforward to show that the
following choices work as desired:
\begin{align}
\zeta=0 &\colon
\begin{cases}
\cJ_1(z) = 1 , \\
\cJ_2(z) = (1-z)^{\min(\beta,1-2\epsilon)} , \\
\cJ_3(z) = (1-z)^{\min(\beta,2-4\epsilon)} ,
\end{cases}
\\
\zeta\ne0 &\colon
\begin{cases}
\cJ_1(z) = 1 , \\
\cJ_2(z) = (1-z)^{1-\beta-\epsilon} , \\
\cJ_3(z) = (1-z)^{1-\beta-\epsilon} ,
\end{cases}
\end{align}
where, for $\zeta\ne0$, we restrict to $\beta \in
(\frac{1}{2}, 1)$ and we choose $\epsilon < 1-\beta$.

\section{Check of hypothesis (S2)}
\label{apx:S2}

In this appendix we show that the hypotesis (S2) of Appendix~\ref{apx:deltaexpansion}
is verified, namely that the spectrum of $S_{\tomega}$ consists only of $\mathbb{R}$
together with at most two isolated points in $\mathbb{C}\setminus\mathbb{R}$, 
symmetric with respect to complex conjugation.

The Green's distribution $\mathcal{G}_{\tilde{\omega},\zeta}$ computed 
in Appendix~\ref{apx:calculation-delta-expansion} has a branch cut at $\Im[\tomega]=0$ and for certain values 
of $\zeta$ it can have poles with $\Im[\tomega] \ne 0$,
which from the explicit calculations of Appendix~\ref{apx:calculation-delta-expansion} coincide with
the zeros of the normalization coefficient
$\mathcal{N}_{\tomega,\zeta}$ in \eqref{eq:Wronski_BC}.

By direct inspection, we know that $\mathcal{N}_{\tomega,\zeta}$ has at most isolated 
zeros, that are reflection symmetric about the real axis. These bound state
frequencies form a set $\BS_\zeta \subset \bC$,  with
$\BS_\zeta = \BS_\zeta^+ \cup \overline{\BS_\zeta^+}$ with
$\Im[\BS_\zeta^+] > 0$. We conclude that $\sigma(S_\tomega) =
\mathbb{R} \cup \BS_\zeta$.
By general arguments from Appendix~\ref{apx:deltaexpansion}, the resolvent
$T_{\tomega}=S^{-1}_{\tomega}$ is analytic on its resolvent
set $\rho(S_\tomega) = \bC \setminus \sigma(S_\tomega)$. 

We will now argue that either $\BS_\zeta^+ = \varnothing$ or
$\BS_\zeta^+ = \{ \tomega_\zeta \}$ consists of a single point. 
Using the notation from
Appendix~\ref{apx:calculation-delta-expansion}, the zeros of $\mathcal{N}_{\tomega,\zeta}$ are precisely the solutions of the
transcendental equation
\begin{equation} \label{eq:BS-zeta}
\tan(\zeta) = \frac{B}{A} \doteq \Theta(\tomega)
\end{equation}
in the upper half complex plane, $\Im[\tomega] > 0$ and $\zeta \in
[0,\pi)$, together with their complex conjugates. $A$ and $B$ are as in \eqref{eq:A_B_constants}.
When $\zeta = \pi/2$, we interpret any $\tomega$ at which $\Theta(\tomega)$ has a pole as a
solution of~\eqref{eq:BS-zeta}. When written out explicitly, the
RHS of~\eqref{eq:BS-zeta} is a ratio of products of
gamma functions with $\tomega$-dependent parameters. Its main
characteristics are that, for generic values of the parameters, it has only the simple zeros at $\tomega_\pm(n)$
and the simple poles at $\tomega^\pm(n)$ for
$n=0,1,2,\ldots$, where
\begin{align}
\label{eq:zeros}
\tomega_\pm(n)
&= \pm \frac{k}{\ell} -k\Omega_\H 
- 2i(n+\beta) \frac{(r_+ \mp r_-)}{\ell^2} \ ,
\\
\label{eq:poles}
\tomega^\pm(n)
&= \pm \frac{k}{\ell} -k\Omega_\H 
- 2i(n+1-\beta) \frac{(r_+ \mp r_-)}{\ell^2} \ ,
\end{align}
as well as the asymptotic behavior 
\begin{multline} \label{eq:asymptTheta}
\Theta(\tomega)
= \frac{\Gamma(\sqrt{\mu^2+1})}{\Gamma({-\sqrt{\mu^2+1}})}
\left(\frac{\ell^4 (-i\tomega)^2}{4(r_+^2-r_-^2)}\right)^{-\sqrt{\mu^2+1}}
\\
\times [1 + \mathcal{O}(|\tomega|^{-1})]
\end{multline}
for $|\tomega| \to \oo$, which follows from the Stirling asymptotic
formula. The branch of the power function must agree with the principal
branch when $-i\tomega > 0$. Some of the poles or zeros may merge for
special values of the parameters.

The zeros and poles of $\Theta(\tomega)$ give us the explicit solutions
of~\eqref{eq:BS-zeta}, respectively, for $\zeta = 0$ (Dirichlet) and
$\zeta = \pi/2$ (Neumann) boundary conditions. For a general value of
$\zeta$, the transcendental nature of equation~\eqref{eq:BS-zeta}
prevents us from giving explicit solutions. Although this equation
could certainly be solved numerically for any value of the parameters $\mu^2$,
$\ell$, $r_+$, $r_-$ and $k$ describing the BTZ black hole
and the scalar field, we can make the following qualitative
conclusions.

Since $\zeta$ is always real, $\tomega \in \bC$ for which
$\Theta(\tomega) \not\in \bR$ is never a solution of~\eqref{eq:BS-zeta}.
On the other hand, when $\Theta(\tomega)$ is real,
equation~\eqref{eq:BS-zeta} is certainly satisfied for $\zeta =
\arctan(\Theta(\tomega))$. Thus, for fixed $\zeta$, the solutions
of~\eqref{eq:BS-zeta} exist and lie on the lines of real phase $\arg
[\Theta(\tomega)] = 0$ or $\pi$. Roughly speaking, lines of real phase
stretch between the poles and zeros of $\Theta(\tomega)$, also with one such
line stretching to $\oo$ through the upper half plane from the
pole with the largest $\Im[\tomega]$, as can be deduced by \eqref{eq:asymptTheta}.

%This qualitative behavior is illustrated in Figures~\ref{fig:phase-mu2-neg} and~\ref{fig:phase-mu2-pos} with
%numerical plots of the phase contours of $\Theta(\tomega)$.

%\begin{figure}
%	\caption{Case $-1 < \mu^2 < 0$: zeros and poles of $\Theta(\tomega)$
%		from~\eqref{eq:BS-zeta}, along with contours of equal $\arg \Theta(\tomega)$.
%		Solid lines denote real phase $\arg \Theta(\tomega) = 0$ or $\pi$.}
%	\label{fig:phase-mu2-neg}
%\end{figure}

%\begin{figure}
%	\caption{Case $\mu^2 > 0$: zeros and poles of $\Theta(\tomega)$
%		from~\eqref{eq:BS-zeta}, along with contours of equal $\arg \Theta(\tomega)$.
%		Solid lines denote real phase $\arg \Theta(\tomega) = 0$ or $\pi$.}
%	\label{fig:phase-mu2-pos}
%\end{figure}

In the case $\mu^2 \geqslant 0$, only the $\zeta = 0$ (Dirichlet)
boundary condition is allowed (see Section~\ref{sec:endpoints}), which
corresponds to zeros of $\Theta(\tomega)$. As it can be seen
from~\eqref{eq:zeros}, all of the zeros are confined to ther lower half complex plane and
so there are no solutions of~\eqref{eq:BS-zeta} with $\Im[\tomega] > 0$. 
Therefore, in this case, there are no bound state frequencies, $\BS_\zeta^+ =
\varnothing$.

When $-1 < \mu^2 < 0$, all the poles and zeros lie in the lower half complex plane and closest to the
real axis is the pole at
\[
\tomega^+(0)
= \frac{k}{\ell} - k \Omega_\H
-i\left(1-\sqrt{\mu^2+1}\right) \frac{(r_+-r_-)}{\ell^2} \ .
\]
%In the case $-1 < \mu^2 < 0$, Figure~\ref{fig:phase-mu2-neg} indicates that
The solutions with $\Im[\tomega] > 0$ must lie on the single line
of real phase stretching from this pole and are
parametrized by $\zeta \in [\zeta_*,\pi)$. This phase line crosses the
$\Im[\tomega] = 0$ line at $\tomega = 0$, where
\begin{equation*}
\Theta(0)
= \frac{\Gamma\left(2\beta-1\right)\left|\Gamma\left(1-\beta+i\ell\frac{k}{r_+}\right)\right|^2}{\Gamma\left(1-2\beta\right)\left|\Gamma\left(\beta+i\ell\frac{k}{r_+}\right)\right|^2}
= \tan(\zeta_*) \ .
\end{equation*}
Since $\beta \in (\frac{1}{2},1)$, then 
$\zeta_* \in (\frac{\pi}{2}, \pi)$. Qualitatively, we also
see that the solution $\tomega = \tomega_\zeta$ is simple%
\footnote{This could be rigorously established by a careful
	application of the \emph{argument principle}, which we omit for
	brevity, to the function $f(\tomega) = \tan(\zeta) - \Theta(\tomega)$,
	which confirms the existence of a single simple zero $\tomega_\zeta \in
	\Im[\tomega]$ provided the integrals $\oint
	\frac{f'(\tomega)}{f(\tomega)} \frac{\dd \tomega}{2\pi i}$ stabilize
	to the value $1$ over a sequence of simple closed and positive
	contours whose interior exhausts the upper half complex plane.}
and of course isolated. Hence, in this case $\BS_\zeta = \{\tomega_\zeta ,
\overline{\tomega_\zeta} \}$. The real and imaginary parts of
$\tomega_\zeta$ are plotted as a function of $\zeta$ in
Figure~\ref{fig:bound-state} for a particular value of other parameters.

\section{Check of hypothesis (S3)}
\label{apx:S3}

In this appendix we show that the hypothesis (S3) of Appendix~\ref{apx:deltaexpansion}
is verified, namely that there exists a spectral measure for the linearized pencil $\bS_\tomega$ in \eqref{eq:linearpencil}.

Following the notation of Appendix~\ref{apx:deltaexpansion}, the inner product
space $\cK = (\H^2, [-,-])$, with bounded bilinear form $[\textbf{v},
\textbf{u}] = (\textbf{v}, \R \textbf{u})$, defines a \emph{Krein
	space}~\cite{bognar,Langer:1982}, that is, a Banach (in this case
Hilbert) space with a bounded hermitian scalar product that need not be
positive definite. The spectral problem of the linear operator pencil
$\bS_\tomega = \P + \tomega \R$ is equivalent to the standard spectral
problem $-\R^{-1} \P = \tomega \mathbf{I}$, where the operator $\A
\doteq -\R^{-1} \P$ is now self-adjoint with respect to the Krein space
scalar product $[-,-]$.

Unfortunately, unlike the Hilbert space case, there is no spectral
theorem available for an arbitrary self-adjoint operator on a Krein
space. However, there are some special cases where the spectral theorem,
and hence the existence of a spectral measure $\E(\nu)$ as requested by
hypothesis (S3) in Appendix~\ref{apx:deltaexpansion}, is available. One such
case is when $\A$ is \emph{definitizable}, that is, when there exists
a degree $k$ polynomial $p(\tomega)$ with real coefficients such that
$[\mathbf{u}, p(\A)\mathbf{u}] \geqslant 0$ for each $\mathbf{u} \in
D(\A^k)$. The corresponding spectral theorem can be found
in~\cite{Langer:1982} and~\cite{KaltenbaeckPruckner:2015}. Below, we
give a brief argument verifying that the operator $\A$ discussed in
Appendices~\ref{apx:deltaexpansion} and~\ref{apx:calculation-delta-expansion} is definitizable, hence
fulfilling hypothesis (S3).

The argument is as follows. First, suppose that there exists a definitizable closed
restriction $\A_0$ of $\A$ to a smaller domain $D(\A_0) \subset D(\A)$,
since $[\mathbf{u}, (-\A_0) \mathbf{u}] \geqslant 0$ for all $\mathbf{u}
\in D(\A_0)$. While $\A_0$ itself may no longer be self-adjoint, the
Krein space analog of the Friedrichs extension~\cite{Curgus:1989} then
gives us a self-adjoint extension $\A_1$ that is still satisfies
$[\mathbf{u}, (-\A_1) \mathbf{u}]$ on its domain. Second, since $\A$ is
essentially defined by an ordinary differential operator, the difference of the resolvents
\begin{equation} \label{eq:finite-rank}
(\A_1 - \tomega \mathbf{I})^{-1} - (\A - \tomega \mathbf{I})
\end{equation}
is an operator of finite rank, which is described by the so-called
\emph{Krein resolvent formula}~\cite[\textsection106]{AkhiezerGlazman}.
The finiteness of the rank comes from the fact that an ordinary
differential operator has a finite dimensional space of solutions.
Finally, it is also known that when at least one of the Krein
self-adjoint operators $\A_1$ or $\A$ is definitizable and the
difference of their resolvents~\eqref{eq:finite-rank} has finite rank
for at least one $\tomega$ common to both resolvent sets, then both
operators are definitizable~\cite{JonasLanger:1979}.

Recall that we are working with $\H = L^2((0,1); \cJ(z)\, \dd{z})$ and
consider $\mathbf{u} = [\begin{matrix} \Psi & \Phi
\end{matrix}]^T \in D(\A_0)$ consisting of smooth functions with
compact support. Unwinding all the definitions from
Appendices~\ref{apx:deltaexpansion} and~\ref{apx:calculation-delta-expansion}, and writing out
$[\mathbf{u}, (-\A) \mathbf{u}]$ explicitly and using integration by
parts, we get
\begin{multline*}
	(\Psi, (-\cJ^{-1} L_{\tomega=0}) \Psi) + (\Phi, \cR_2 \Phi)
	= \int_0^1 \dd{z} \left[z \left|\frac{\dd \Psi(z)}{\dd z}\right|^2 \right. 
	\\
	\left. + \left(\frac{\ell^2 k^2}{4 r_+^2} + \frac{\mu^2}{4(1-z)}\right) |\Psi(z)|^2
	+ \frac{\ell^4 \cJ(z) |\Phi(z)|^2}{4 (r_+^2-r_-^2)}
	\right]
	\, . \label{eq:integrand}
\end{multline*}
When $\mu^2\geqslant 0$, all the terms appearing under the integral are
manifestly non-negative, meaning that so is the whole integral. When
$-1 < \mu^2 < 0$, the integrand is still non-negative. 
This can be proven observing that, being the term proportional to $k^2$ in the integrand strictly greater than $0$, it suffices to show positivity for $-\mathcal{J}^{-1}L_{\tomega=0}$ when $k=0$. Yet, in this case, in view of \eqref{eq:B1} and of the results of Appendix \ref{apx:deltaexpansion}, $-\mathcal{J}^{-1}L_{\tomega=0}$ is a self-adjoint operator with strictly positive spectrum, which is tantamount to saying that it is a positive operator.
Thus, the restriction of $\A$ to $D(\A_0)$ does satisfy $[\mathbf{u}, (-\A_0)
\mathbf{u}] \geqslant 0$ for all $\mathbf{u} \in D(\A_0)$. By the
preceding reasoning, this finally implies that $\A$ is definitizable.

% BIBLIOGRAPHY

\end{document}